\documentclass[11pt]{article}

\usepackage[final]{acl}

\usepackage{times}
\usepackage{latexsym}

\usepackage[T1]{fontenc}

\usepackage[utf8]{inputenc}

\usepackage{microtype}

\usepackage{inconsolata}

\usepackage{graphicx}

\usepackage{booktabs}

\usepackage[table]{xcolor}
\usepackage[most]{tcolorbox}
\usepackage{fvextra}
\usepackage{caption}
\usepackage{kotex}

\usepackage{array}
\usepackage{tabularx}
\usepackage{makecell}

\usepackage{listings}
\tcbuselibrary{listings,breakable}

\usepackage{amssymb}

\title{KoViDoRe: A Benchmark for Korean Visual Document Retrieval}

\author{
  \textbf{Yongbin Choi}, 
  \textbf{Yongwoo Song}, 
  \textbf{Mujeen Sung}\thanks{Corresponding author} \\
  Kyung Hee University \\
  \texttt{\{yongbinchoi, syw5141, mujeensung\}@khu.ac.kr} \\[0.6em]
  \hspace*{-0.5cm}\includegraphics[height=1em]{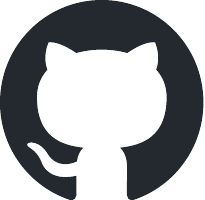}\ 
  \textbf{Code:} \texttt{\url{https://github.com/whybe-choi/kovidore-benchmark}} \\
  \hspace*{-0.5cm}\includegraphics[height=1em]{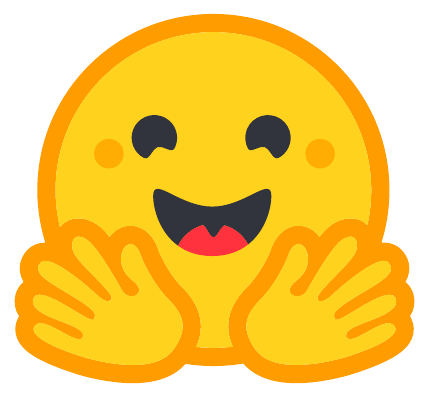}\ 
  \textbf{Dataset:} \texttt{\url{https://hf.co/datasets/NomaDamas/ko-vdr-train-public}} \\
}

\begin{document}
\maketitle
\begin{abstract}

Recent advances in multimodal retrieval have improved the ability to retrieve information from visually rich documents such as PDFs and reports. However, existing benchmarks remain largely centered on English and provide limited coverage of Korean visual documents with complex structures.
Furthermore, most existing Korean resources primarily evaluate single-page retrieval, failing to capture realistic scenarios that require evidence aggregation across multiple pages.
To address these gaps, we introduce \textbf{KoViDoRe}, a benchmark for Korean visual document retrieval. The dataset is constructed from publicly available Korean documents with diverse layouts, including tables, figures, and multi-column structures. We develop a multi-stage data curation pipeline consisting of structured document parsing, synthetic query generation using both summary-based and context-based strategies, and relevance mapping with human verification.
Using KoViDoRe, we evaluate a wide range of multimodal retrieval models and observe that current models struggle to effectively handle Korean visual document retrieval, particularly in settings involving structured content and diverse query types.
Motivated by this finding, we further curate a large-scale training dataset, \textbf{Ko-VDR Train Public}, to support the development of retrieval models tailored to Korean visual documents. Together, KoViDoRe and Ko-VDR Train Public provide a unified benchmark and training resource for Korean visual document retrieval.

\end{abstract}

\section{Introduction}

\begin{figure}[t]
    \centering
    \includegraphics[width=\linewidth]{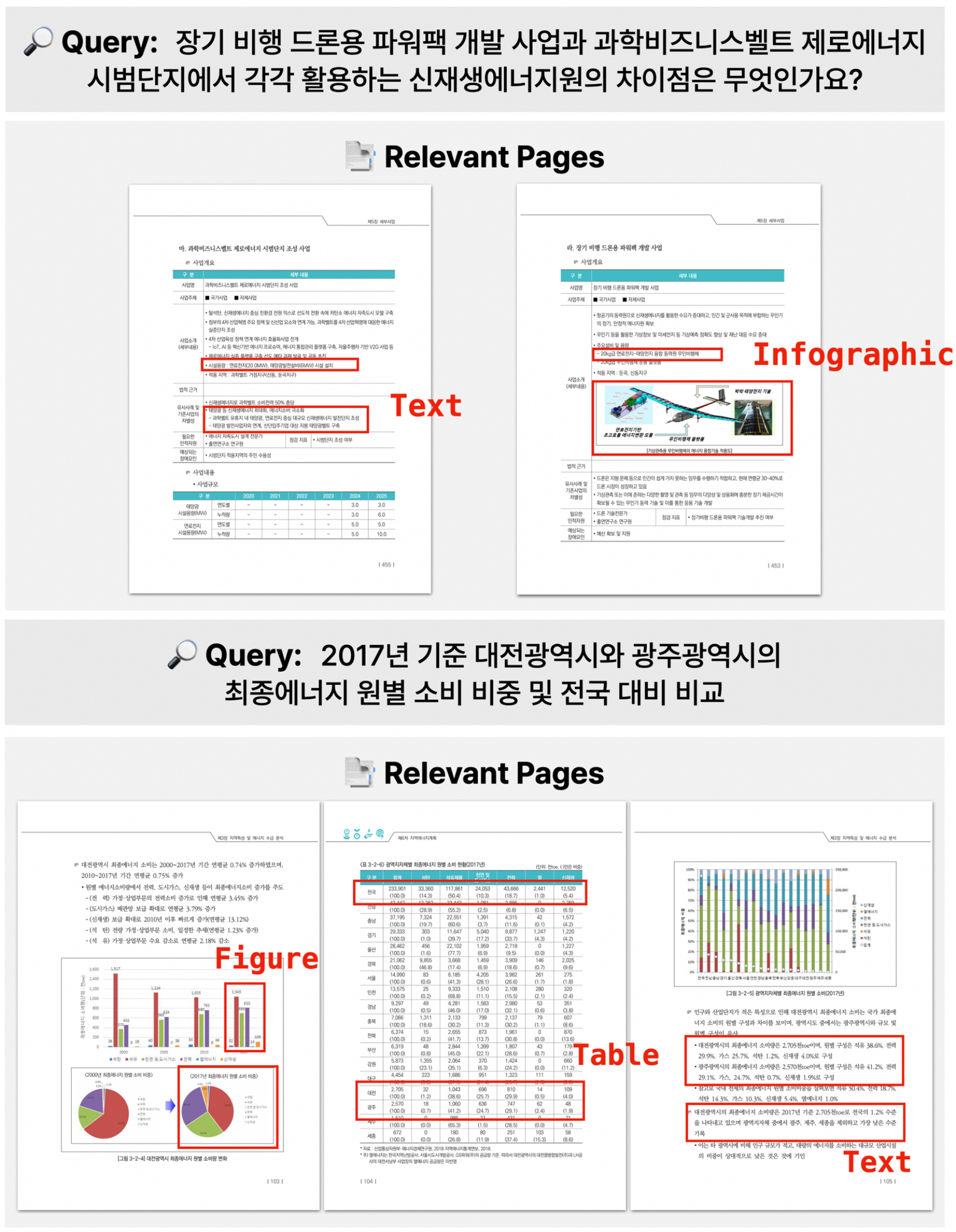}
    \caption{Examples of queries in KoViDoRe and their corresponding relevant pages. Each query requires aggregating evidence from pages and diverse modalities, including text, tables, figures, and infographics.}
    \label{fig:sample}
\end{figure}

Recent advances in multimodal large language models and retrieval-augmented generation (RAG) have significantly improved the ability to retrieve and reason over complex documents \citep{abootorabi-etal-2025-ask, song2025bridge, yu2024visrag}. In particular, a growing line of work on visual document retrieval (VDR) and multimodal retrieval models, including approaches such as ColPali \citep{faysse2024colpali}, has demonstrated strong performance in retrieving document pages by jointly modeling textual, visual, and layout information. These developments have enabled systems to move beyond text-only retrieval and better handle structured documents such as PDFs, reports, and forms \citep{yan2026unlocking}. To support this progress, recent benchmarks have adopted large-scale synthetic data generation pipelines, exemplified by frameworks such as ViDoRe \citep{mace2025vidore, loison2026vidore}, enabling scalable evaluation of multimodal retrieval systems. In parallel, efforts such as Jina-VDR \citep{gunther2025jina}, MIRACL-VISION \citep{osmulski2025miracl} and SDS KoPub VDR \citep{lee2025sds} have extended this paradigm to non-English settings, providing valuable resources for Korean document retrieval and highlighting the importance of multilingual evaluation. 

Despite these advances, existing benchmarks largely formulate VDR as a single-page retrieval task, where each page is treated as an independent unit \citep{wasserman-etal-2025-real, wang2025vidorag}. 
While this formulation simplifies evaluation, it does not accurately reflect how information is organized in real-world documents. 
In practical scenarios such as financial reports, policy documents, and technical manuals, relevant information is often distributed across multiple pages, requiring systems to aggregate evidence and perform reasoning over a set of pages rather than retrieving a single relevant page \citep{cho2024m3docrag}. Although prior datasets may include queries that involve reasoning, such reasoning is typically confined to a single page or limited contextual scopes \citep{dong-etal-2025-mmdocir}. 
Addressing this limitation requires a shift from single-page retrieval to multi-page evidence aggregation, where retrieval systems must identify a coherent set of pages that collectively fulfill the information need.

In this work, we introduce \textbf{KoViDoRe}, a benchmark for Korean visual document retrieval that explicitly models this setting. Building upon prior synthetic data generation approaches, we construct a dataset of realistic enterprise-style documents and generate queries that require  retrieving and synthesizing information distributed across multiple pages to provide a complete answer. Unlike existing benchmarks that primarily focus on single-page retrieval with extractive queries, we formulate the task to address multi-page relevance. In this setting, a single query often corresponds to multiple supporting pages, requiring models to identify the full set of pages whose combined content is necessary to satisfy the information need. To enable scalable construction of such queries, we adopt and adapt synthetic query generation techniques to the Korean document domain. Our pipeline leverages large language models to generate queries with diverse reasoning patterns, including multi-hop inference, numerical comparison, and cross-sectional aggregation, while maintaining explicit mappings between queries and their supporting pages. Rather than introducing a fundamentally new generation method, our focus is on restructuring the task and dataset to better reflect realistic retrieval scenarios, particularly in non-English and enterprise contexts.

In addition, we release \textbf{Ko-VDR Train Public}, a large-scale training dataset aligned with the proposed task, providing a foundation for developing and evaluating retrieval models in Korean multimodal settings. Through extensive experiments, we show that existing multimodal retrieval approaches struggle significantly under this formulation, especially as the number of required supporting pages increases. These results highlight the limitations of current single-page retrieval paradigms and underscore the need for models that can effectively aggregate over evidence distributed across multiple pages. Our contributions are as follows:
\begin{itemize}
    \item We introduce \textbf{KoViDoRe}, a Korean-focused benchmark designed to evaluate multi-page retrieval performance in realistic document settings.
    \item We adapt synthetic query generation techniques to construct realistic queries with explicit page-level supervision.
    \item We release \textbf{Ko-VDR Train Public}, a large-scale dataset supporting training in Korean multimodal retrieval.
    \item We show that existing retrieval models struggle to retrieve evidence distributed across multiple pages on Korean visual documents, highlighting a gap not captured by existing benchmarks.
\end{itemize}

\section{Related Work}

\subsection{Multimodal Retrieval Models}

Recent advances in multimodal retrieval models have significantly improved the ability to retrieve information from visually rich documents \citep{gunther2025jina, ma-etal-2024-unifying, qwen3vlembedding, moreira2026nemotron}. In particular, models such as ColPali and related late-interaction architectures represent each document page as a unified retrieval unit, encoding the textual content, visual features, and layout structure contained within the page \citep{faysse2024colpali, xiao2025metaembed}. This allows retrieval models to capture not only semantic information from text, but also spatial and visual cues, leading to more effective retrieval over visually rich documents. These approaches have demonstrated strong performance across a variety of document understanding tasks, especially in settings where visual structure plays a critical role. In addition to late-interaction models, dual-encoder and dense retrieval approaches have also been extended to multimodal settings, often leveraging vision-language models to capture both textual and visual semantics \citep{ma-etal-2024-unifying, nomicembedmultimodal2025}. These developments have contributed to substantial progress in retrieving relevant content from structured documents such as PDFs, forms, and reports.

However, despite these modeling advances, the datasets used to train and evaluate such models are largely concentrated on English and European languages \citep{yu2024visrag, gunther2025jina, loison2026vidore, peng2025unidoc, wasserman-etal-2025-real, shorten2026irpapers}. As a result, the performance and behavior of multimodal retrieval models on other languages, including Korean, remain underexplored. This is particularly important in document retrieval settings, where linguistic characteristics such as morphology, spacing variation, and domain-specific expressions interact with visual structure and layout. The lack of dedicated Korean benchmarks limits the ability to assess and develop retrieval models for realistic Korean document scenarios.

\subsection{Vision Document Retrieval Benchmarks}

Recent benchmarks for visual document retrieval and document-centric multimodal retrieval, such as ViDoRe \citep{mace2025vidore, loison2026vidore}, Jina-VDR \citep{gunther2025jina}, REAL-MM-RAG \citep{wasserman-etal-2025-real}, UniDoc-Bench \citep{peng2025unidoc},  MIRACL-VISION \citep{osmulski2025miracl}, and IRPAPERS \citep{shorten2026irpapers} have significantly expanded evaluation settings by incorporating visually rich documents, multimodal signals, and realistic query formulations. Many of these benchmarks also support multilingual evaluation. However, their language coverage is largely centered on English and European languages, leaving Korean relatively underrepresented despite its distinct linguistic and document characteristics.

Jina-VDR, for example, includes a Korean subset and broadens the diversity of visual documents and query types. However, its document collections are constructed to cover a wide range of modalities and scenarios, which can make them less representative of real-world Korean document retrieval settings, such as structured public documents, reports, or enterprise-style materials. Similarly, MIRACL-VISION provides multilingual evaluation with Korean queries and documents, but its corpus is primarily derived from Wikipedia, which differs substantially from the types of structured and visually complex documents commonly encountered in real-world Korean retrieval scenarios. This discrepancy limits its ability to fully capture the challenges of practical document retrieval in Korean contexts. SDS KoPub VDR \citep{lee2025sds} addresses this gap by introducing a large-scale benchmark for Korean visual document retrieval. It provides an important step toward evaluating retrieval models on Korean documents with complex layouts and diverse structures. Nevertheless, its query formulation remains strictly focused on single-page retrieval, where each query is mapped to only one relevant page, rather than addressing information distributed across multiple pages.

In contrast, KoViDoRe is designed to emphasize more complex information needs that cannot be satisfied by a single page alone. Our queries require aggregating evidence across multiple pages and capturing relationships between distributed pieces of information. By focusing on Korean documents while introducing queries with higher reasoning complexity and more realistic document distributions, KoViDoRe complements existing benchmarks and provides a more challenging evaluation setting for Korean visual document retrieval.

\begin{figure*}[t]
    \centering
    \includegraphics[width=\textwidth]{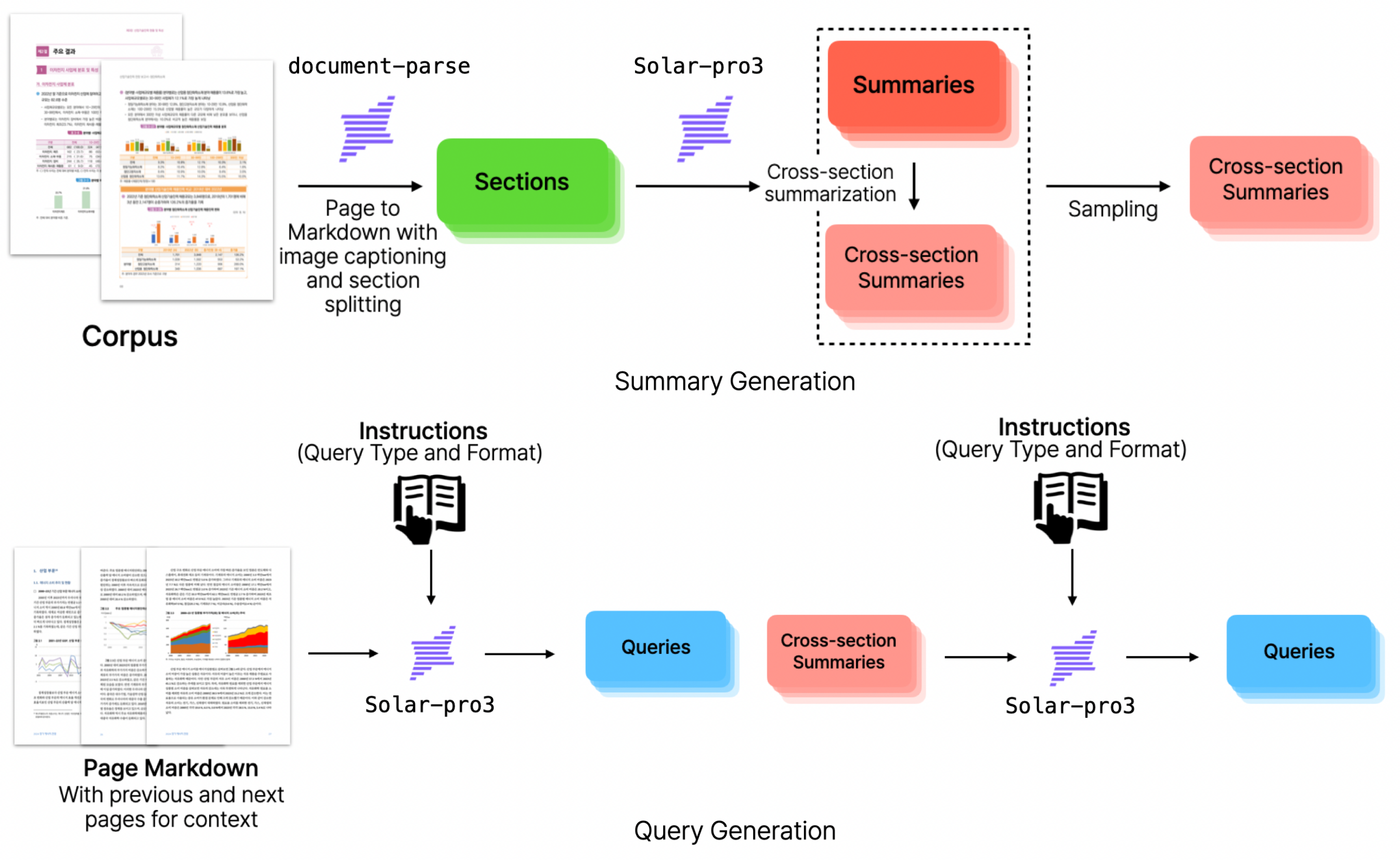}
    \caption{Overview of the KoViDoRe data generation pipeline. Queries are generated through both summary-based and context-based strategies. Summary representations capture global document relationships, while context-based generation focuses on local content. The pipeline further includes filtering and manual verification to ensure query quality and reliable relevance mapping.}
    \label{fig:data_pipeline}
\end{figure*}

\section{Dataset Curation}
\label{sec:data_curation}

As illustrated in Figure~\ref{fig:data_pipeline}, our dataset curation pipeline consists of document collection, structured parsing, query generation, and relevance mapping. The overall pipeline design is inspired by ViDoRe V3 \citep{loison2026vidore}, and is adapted to better reflect the characteristics of Korean document ecosystems and real-world data sources.

\subsection{Source Collection}

To construct a realistic benchmark for Korean visual document retrieval, we collect document corpora from publicly available sources, including government reports, policy documents, and enterprise-style materials obtained from the Korean public data portal\footnote{https://www.data.go.kr} and official institutional websites. We prioritize documents that are freely available under the Korea Open Government License (KOGL) Type 1 and 2, as well as materials without restrictive licensing conditions. This ensures that the collected data can be used for research and downstream applications without legal constraints. Following SDS KoPub VDR \citep{lee2025sds}, which also leverages publicly accessible Korean documents, we focus on authentic document sources rather than synthetic or simplified formats. This allows the benchmark to reflect real-world document characteristics, including complex layouts, tables, figures, and multi-column structures.

\subsection{Document Processing and Parsing}

We first split each PDF document into individual pages and perform processing at the page level. This allows us to treat each page as a fundamental unit while preserving the document structure. To extract structured representations from each page, we employ the Upstage Document Parse\footnote{document-parse-251217}, which is effective for parsing structurally complex Korean documents while preserving layout-aware information. For every page, textual content and visual elements are identified and organized into semantically meaningful sections. Specifically, the parser provides both page-level markdown and element-level markdown for each page. The page-level markdown offers a unified view of the entire page content, while the element-level markdown decomposes the page into fine-grained components such as text blocks, tables, figures, charts, and diagrams. In addition to structural extraction, the document parser provides captions for visual elements such as figures, charts, and diagrams. These captions offer semantic descriptions of visual content, enabling better understanding of non-textual information during downstream processing. By combining page-level and element-level markdown with captioned visual elements, our preprocessing pipeline preserves document-level structure while enabling fine-grained and semantically enriched access to page content.

\subsection{Query Generation}

To construct a diverse and scalable set of queries, we adopt a synthetic query generation pipeline based on reasoning-oriented language model, Solar-Pro3\footnote{solar-pro3-260126}, which supports strong Korean language understanding and is well-suited for generating complex Korean queries that reflect multiple information needs. The model generates queries by conditioning on document content, including both textual and visual information extracted during preprocessing.

\paragraph{Summary-based Query Generation}

Before query generation, we first construct intermediate summaries to better expose document structure and cross-page relationships. Specifically, we generate two types of summaries. The first type consists of single-section summaries that describe individual sections. The second type consists of cross-section summaries, which are constructed by randomly sampling multiple single-section summaries (e.g., 3, 5, or 7 sections) and synthesizing them to capture cross-sectional relationships. These summaries provide a higher-level abstraction of document content, allowing the generation process to capture relationships that are not easily observable from isolated pages. Queries generated from summaries therefore tend to reflect more global information needs and often require reasoning across multiple sections or pages.

\paragraph{Context-based Query Generation}

In addition to summary-based generation, we also generate queries directly from local document context. In this setting, the model is prompted using page-level or local multi-page context windows, enabling it to produce queries grounded in nearby content. This complementary route helps capture more localized information needs and preserves natural query patterns tied to specific document regions. Combining both summary-based and context-based queries allows the dataset to cover a broader spectrum of retrieval scenarios.

\paragraph{Diversity Control}

To promote diversity, we adopt the query formulation scheme introduced in ViDoRe V3 \citep{loison2026vidore}. Specifically, we control the query format and type during generation, enabling the construction of a diverse set of queries with different structural patterns and information needs. As a result, the dataset includes various query types such as multi-hop reasoning, numerical comparison, and aggregation-based queries, where a single query may exhibit multiple types simultaneously. The full set of query types and formats, along with their definitions, is summarized in Table~\ref{tab:query-type-definitions} and Table~\ref{tab:query-format-definitions}.

\begin{table*}[t]
    \centering
    \small
    \begin{tabular}{lccccc}
        \toprule
        \textbf{Subset} & \textbf{\#Docs} & \textbf{\#Pages} & \textbf{\#Queries} & \textbf{\#Qrels} & \textbf{Avg.\ Pages / Query} \\
        \midrule
        Cybersecurity & 17 & 1,150 & 149 & 409 & 2.74 \\
        HR            & 9  & 2,109 & 221 & 726 & 3.30 \\
        Energy        & 11 & 1,993 & 173 & 525 & 3.03 \\
        Economic      & 20 & 1,477 & 163 & 413 & 2.55 \\
        \midrule
        Total         & 57 & 6,729 & 706 & 2,073 & 2.94 \\
        \bottomrule
    \end{tabular}
    \caption{Statistics of the KoViDoRe benchmark across domain-specific subsets. Avg.\ Pages / Query denotes the average number of relevant pages per query.}
    \label{tab:kovidore_stats}
\end{table*}

\begin{figure}[t]
    \centering
    \includegraphics[width=\linewidth]{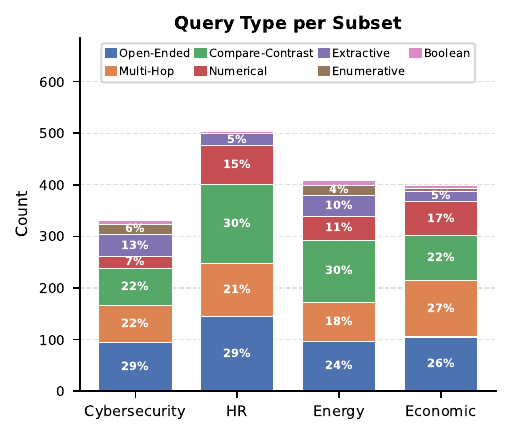}
    \caption{Distribution of query types across subsets. Note that query types are not mutually exclusive, accounting for the multi-faceted nature of complex information needs.}
    \label{fig:query_type_stats}
\end{figure}

\begin{figure}[t]
    \centering
    \includegraphics[width=\linewidth]{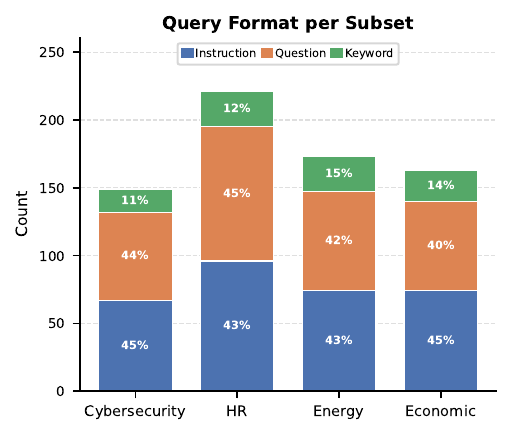}
    \caption{Distribution of query formats across subsets.}
    \label{fig:query_format_stats}
\end{figure}

\subsection{Relevance Mapping and Filtering}

To construct reliable query-document relevance annotations, we incorporate relevance mapping into multiple stages of the pipeline. During query generation, the model is provided with document content in markdown format, including both page-level and element-level representations, and queries are generated based on specific sections or combinations of sections, implicitly capturing initial relevance signals. After query generation, we perform an additional relevance mapping step at the page level by evaluating each query against candidate document pages to identify supporting evidence.

To improve annotation quality while reducing manual effort, we apply both consistency-based and rule-based filtering. Relevance signals from the generation stage are compared with those from the additional mapping stage, and only consistent pairs are retained. While this process may discard some challenging cases where relevant pages are difficult to identify during mapping, we prioritize annotation reliability over coverage, as relevance judgments can be inherently ambiguous in multimodal and multi-page settings. In addition, we remove low-quality queries such as those with excessive keyword enumeration or those that directly reveal answer content from the document, resulting in a reduced but more reliable set of keyword-based queries. For the benchmark, we further incorporate human verification, where annotators perform a final review of query-page relevance annotations and refine queries through rephrasing when they are unnatural or ambiguous. This process balances automatic filtering and human verification to produce high-quality annotations.

\subsection{Dataset Statistics}

Table~\ref{tab:kovidore_stats} summarizes the overall statistics of the benchmark across its four domain-specific subsets. Figure~\ref{fig:query_type_stats} and Figure~\ref{fig:query_format_stats} further illustrate the distributions of query types and query formats. The benchmark consists of 57 documents and 6,729 pages, with a total of 706 queries and 2,073 relevance annotations. The four domain-specific subsets exhibit notable differences in scale and structure. For example, the HR and Energy subsets contain a larger number of pages per query, suggesting that queries in these domains often require aggregating information from multiple pages. As shown in Figure~\ref{fig:query_type_stats} and Figure~\ref{fig:query_format_stats}, the dataset contains a diverse set of query types and formats across all subsets. Multi-hop, open-ended, and comparison-based queries appear frequently, while question- and instruction-style queries are more common than keyword queries.

\section{Experiments}

\subsection{Experimental Setup}

We evaluate Korean visual document retrieval as a ranking task over document pages. Given a query $q$ and a collection of document pages $\mathcal{D}$, the goal is to retrieve and rank pages that are relevant to the query. Each document is represented as a set of pages containing both textual and visual content. Queries are written in Korean and reflect diverse information needs grounded in real-world documents. Relevance is defined at the page level with graded labels: a page is labeled as fully relevant (2) if it contains sufficient information to answer the query, and partially relevant (1) if it provides supporting evidence. The dataset follows the BEIR-style evaluation framework with a corpus, queries, and relevance judgments (qrels) \citep{thakur2021beir}. We report performance using nDCG@10. We evaluate a range of multimodal retrieval models on the KoViDoRe benchmark, covering small ($<$1B), medium (1B--4B), and large ($\geq$4B) models. The evaluated models include CLIP-based encoders \citep{radford2021learning, zhai2023sigmoid, koukounas2024jina}, late-interaction retrieval models such as ColPali and ColQwen \citep{faysse2024colpali, nomicembedmultimodal2025, huang2025beyond}, and recent multimodal embedding models \citep{jiang2024vlm2vec, gunther2025jina, qwen3vlembedding}. Evaluation is conducted across four domains: Cybersecurity, Energy, Economic, and Human Resources.

We conduct evaluation using the MTEB (Massive Text Embedding Benchmark) \citep{muennighoff2022mteb}, adapted to support multimodal retrieval on the KoViDoRe dataset. This provides a standardized and reproducible evaluation pipeline across all models. We build separate retrieval indices for each domain subset and each query is evaluated only against the pages within its corresponding domain. 

\begin{table*}[t]
\centering
\small
\begin{tabular}{lcccccc}
\toprule
\textbf{Model} & \textbf{Params} & \textbf{Cyber} & \textbf{Energy} & \textbf{Economic} & \textbf{HR} & \textbf{Avg.} \\
\midrule

\rowcolor{gray!15}
\multicolumn{7}{c}{\textit{Small Models ($<$1B parameters)}} \\
\midrule
openai/clip-vit-base-patch16$^{\spadesuit}$ & 151M & 4.1 & 0.8 & 0.0 & 0.6 & 1.4 \\
vidore/colSmol-256M$^{\blacklozenge}$ & 256M & 19.7 & 9.5 & 1.0 & 1.1 & 7.8 \\
vidore/colSmol-500M$^{\blacklozenge}$ & 500M & 26.2 & 9.9 & 0.6 & 0.9 & 9.4 \\
jinaai/jina-clip-v2$^{\spadesuit}$ & 865M & 20.4 & 11.3 & 0.2 & 3.1 & 8.8 \\
google/siglip-so400m-patch14-384$^{\spadesuit}$ & 878M & 15.3 & 5.3 & 1.3 & 1.1 & 5.8 \\

\midrule
\rowcolor{gray!15}
\multicolumn{7}{c}{\textit{Medium Models (1B--4B parameters)}} \\
\midrule
Qwen/Qwen3-VL-Embedding-2B$^{\clubsuit}$ & 2.0B & 61.3 & 40.6 & 15.3 & 18.7 & 34.0 \\
vidore/colqwen2-v1.0$^{\blacklozenge}$ & 2.2B & 53.3 & 42.0 & 8.0 & 14.7 & 29.5 \\
vidore/colpali-v1.1$^{\blacklozenge}$ & 2.9B & 31.9 & 18.2 & 3.0 & 6.0 & 14.8 \\
vidore/colpali-v1.2$^{\blacklozenge}$ & 2.9B & 33.2 & 16.4 & 2.1 & 4.5 & 14.1 \\
vidore/colpali-v1.3$^{\blacklozenge}$ & 2.9B & 34.7 & 20.6 & 1.6 & 6.2 & 15.8 \\
ApsaraStackMaaS/EvoQwen2.5-VL-Retriever-3B-v1$^{\blacklozenge}$ & 3.0B & 41.4 & 31.5 & 6.3 & 11.3 & 22.6 \\
nomic-ai/colnomic-embed-multimodal-3b$^{\blacklozenge}$ & 3.0B & 47.4 & 44.2 & 10.5 & 32.9 & 33.7 \\
vidore/colqwen2.5-v0.2$^{\blacklozenge}$ & 3.0B & 43.9 & 44.3 & 3.9 & 13.5 & 26.4 \\
jinaai/jina-embeddings-v4$^{\clubsuit}$  & 3.8B & \underline{77.6} & \textbf{67.7} & \textbf{24.5} & \textbf{50.1} & \textbf{55.0} \\

\midrule
\rowcolor{gray!15}
\multicolumn{7}{c}{\textit{Large Models ($\geq$4B parameters)}} \\
\midrule
TomoroAI/tomoro-colqwen3-embed-4b$^{\blacklozenge}$ & 4.0B & 55.3 & 31.0 & 9.1 & 10.1 & 26.4 \\
eagerworks/eager-embed-v1$^{\clubsuit}$  & 4.0B & 51.5 & 32.7 & 5.4 & 7.0 & 24.2 \\
TIGER-Lab/VLM2Vec-Full$^{\clubsuit}$  & 4.2B & 9.8 & 3.2 & 1.3 & 1.3 & 3.9 \\
ApsaraStackMaaS/EvoQwen2.5-VL-Retriever-7B-v1$^{\blacklozenge}$ & 7.0B & 66.0 & 55.4 & 12.1 & 26.4 & 40.0 \\
nomic-ai/colnomic-embed-multimodal-7b$^{\blacklozenge}$ & 7.0B & 69.6 & 59.5 & 12.4 & 33.3 & 43.7 \\
Qwen/Qwen3-VL-Embedding-8B$^{\clubsuit}$ & 8.0B & \textbf{77.8} & \underline{63.2} & \underline{23.4} & \underline{37.4} & \underline{50.4} \\
TomoroAI/tomoro-colqwen3-embed-8b$^{\blacklozenge}$ & 8.0B & 73.7 & 58.5 & 16.3 & 26.5 & 43.8 \\

\bottomrule
\end{tabular}
\caption{
Performance comparison on KoViDoRe benchmark (nDCG@10, \%).  
\textbf{Bold} indicates the best score; \underline{underline} indicates the second-best score. $\spadesuit$: CLIP-based, $\blacklozenge$: late-interaction, $\clubsuit$: single-vector models.
Cyber: Cybersecurity, HR: Human Resources.
}
\label{tab:kovidore-v2-results}
\end{table*}

\subsection{Main Results}

Table~\ref{tab:kovidore-v2-results} presents the performance of all evaluated models on the benchmark. Overall, we observe a clear performance gap between model scales. Small models perform poorly across all domains, often failing to retrieve relevant pages. Medium-scale models show improved performance, particularly those based on late-interaction architectures. Larger models generally achieve higher scores, although performance gains are not uniform across domains. Among all models, \texttt{jina-embeddings-v4} achieves the best overall performance, significantly outperforming other models across all domains. This suggests that retrieval effectiveness is influenced not only by model scale, but also by factors such as training objective and data composition.

\subsection{Analysis}

Despite improvements from larger models, performance remains limited across all domains. In particular, domains such as Economic and Human Resources consistently show lower scores, indicating that retrieving relevant information in these settings is especially challenging. This is likely due to more complex document structures and the presence of information distributed across multiple pages.

We also observe that even strong retrieval models achieve relatively limited performance on several subsets of KoViDoRe. This suggests that the benchmark introduces additional challenges beyond conventional document retrieval settings, including complex layouts, structured visual content, and information distributed across multiple pages. In particular, queries associated with multiple relevant pages remain difficult for existing models, indicating that effectively retrieving and aggregating distributed document evidence is still a challenging problem in Korean visual document retrieval.

Motivated by this limitation, we further investigate whether training on Korean-specific data can improve retrieval performance, which we explore in the following subsection.

\begin{table*}[t]
\centering
\small
\begin{tabular}{lcccccc}
\toprule
\textbf{Model} & \textbf{Params} & \textbf{Cyber} & \textbf{Energy} & \textbf{Economic} & \textbf{HR} & \textbf{Avg.} \\
\midrule
vidore/colSmol-500M & 500M & 26.2 & 9.9 & 0.6 & 0.9 & 9.4 \\
\quad + Ko-VDR Train Public & & 39.4 & 35.0 & 14.4 & 18.7 &  26.9 \\
\midrule
vidore/colqwen2-v1.0 & 2.2B & 53.3 & 42.0 & 8.0 & 14.7 & 29.5 \\
\quad + Ko-VDR Train Public & & 75.6 & \underline{67.6} & 18.3 & \underline{49.6} &  \underline{52.8} \\
\midrule
jinaai/jina-embeddings-v4 & 3.8B & \underline{77.6} & \textbf{67.7} & \textbf{24.5} & \textbf{50.1} & \textbf{55.0} \\
TomoroAI/tomoro-colqwen3-embed-4b & 4.0B & 55.3 & 31.0 & 9.1 & 10.1 & 26.4 \\
Qwen/Qwen3-VL-Embedding-8B & 8.0B & \textbf{77.8} & 63.2 & \underline{23.4} & 37.4 & 50.4 \\
\bottomrule
\end{tabular}
\caption{Comparison with representative retrieval baselines on KoViDoRe (nDCG@10, \%). \textbf{Bold} indicates the best score; \underline{underline} indicates the second-best score. Cyber: Cybersecurity, HR: Human Resources.}
\label{tab:ko_vdr_training_results}
\end{table*}

\subsection{Ko-VDR Train Public}
\label{sec:ko_vdr_train_public}

\paragraph{Dataset Construction}

To address the limitations identified in the previous section, we curate a large-scale training dataset, \textbf{Ko-VDR Train Public}, using the same data generation pipeline. The dataset consists of query-page pairs derived from Korean visual documents and includes a total of \textbf{310,226} query-page pairs. To ensure data quality, we apply both consistency-based and rule-based filtering. The consistency-based filtering retains only query-page pairs where relevance signals from the query generation stage and the additional relevance mapping stage agree. In addition, we apply rule-based filtering to remove low-quality queries, including those with excessive keyword enumeration and those that directly reveal answer content from the document. This helps eliminate trivial or overly extractive cases and improves the robustness of the data. Unlike the benchmark construction process, we do not perform additional human verification for the training dataset to maintain scalability.

\paragraph{Training Setup}

We fine-tune two late-interaction retrieval models, \texttt{colSmol-500M} and \texttt{colqwen2-v1.0}, using the \texttt{colpali\_engine} framework. Training is conducted on $2\times$ NVIDIA B200 GPUs using BF16. We use a batch size of 128 per device and train for 3 epochs. In addition to Ko-VDR Train Public, we mix in a private Korean visual QA dataset containing TableVQA- and FigureVQA-style supervision \citep{kim2024tablevqa, kahou2017figureqa}. This additional dataset complements the retrieval objective by providing stronger supervision for structured visual understanding, particularly for tables and figures.

\paragraph{Results}

Table~\ref{tab:ko_vdr_training_results} shows that fine-tuning on our dataset consistently improves performance across both \texttt{colSmol-500M} and \texttt{colqwen2-v1.0}. The smaller \texttt{colSmol-500M} model achieves substantial gains across all domains, demonstrating the effectiveness of the proposed training data even for lightweight models. For \texttt{colqwen2-v1.0}, fine-tuning leads to significant performance improvements, achieving competitive results with strong multimodal embedding models and surpassing \texttt{Qwen3-VL-Embedding-8B} despite being smaller in scale. These results highlight that training on Korean-specific data can substantially improve retrieval performance and enable smaller models to compete with larger counterparts.

\subsection{Interpretability}

\begin{figure}[t]
    \centering
    \includegraphics[width=0.9\linewidth]{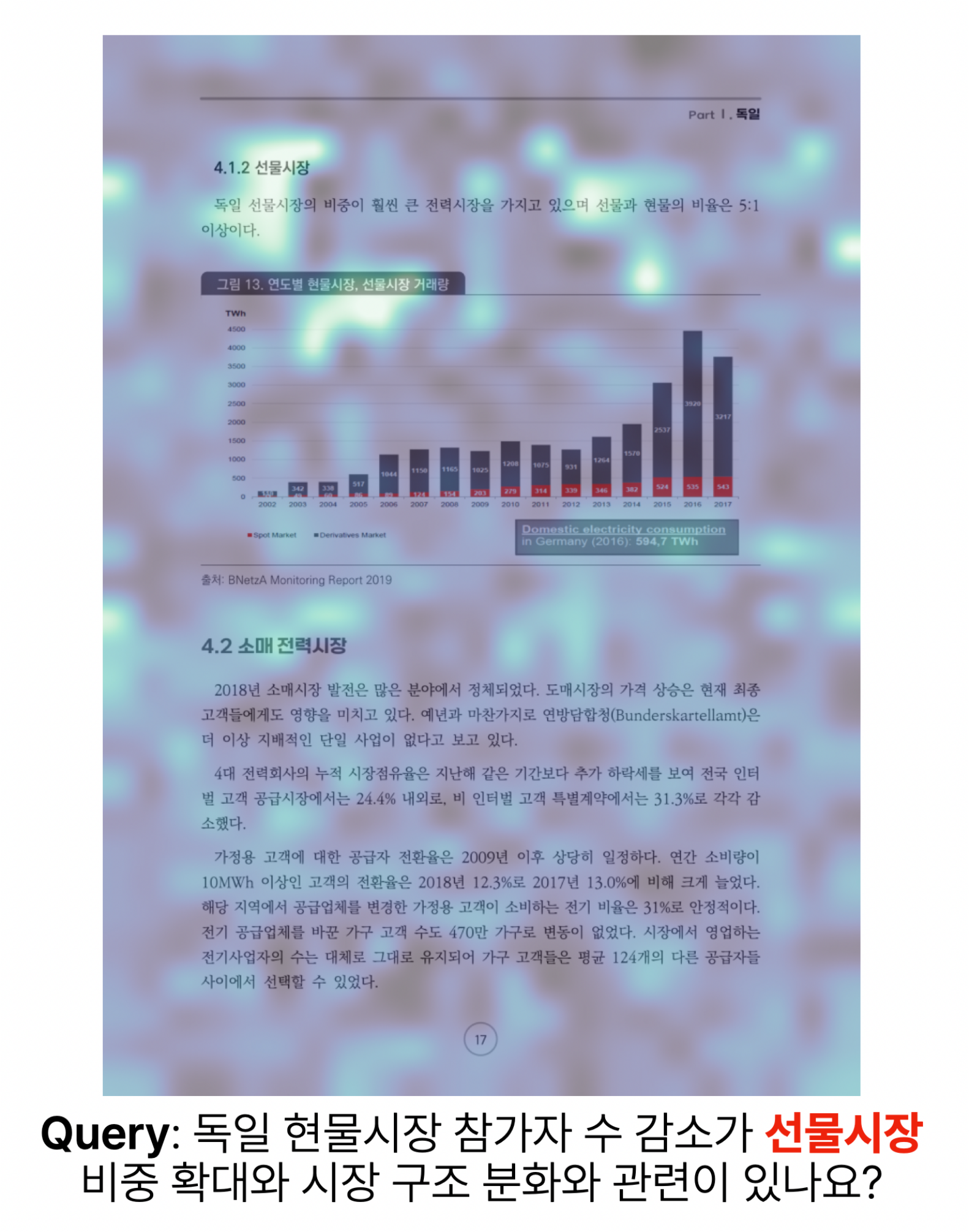}
    \caption{Similarity map on a document example.}
    \label{fig:interpretability}
\end{figure}

To better understand model behavior, we visualize query-to-document similarity using heatmaps for the fine-tuned \texttt{colqwen2-v1.0} model. In Figure~\ref{fig:interpretability}, the model assigns high similarity to the term ``선물시장,'' indicating that it correctly focuses on the key textual evidence relevant to the query. This suggests that the model is able to identify and attend to query-relevant terms in Korean visual documents. Additional examples are provided in Figure~\ref{fig:interpretability_appendix}.

\begin{table*}[t]
\centering
\small
\begin{tabular}{lccccc}
\toprule
\textbf{Model} & \textbf{Cyber} & \textbf{Energy} & \textbf{Economic} & \textbf{HR} & \textbf{Avg.} \\
\midrule

vidore/colqwen2-v1.0 & 53.3 & 42.0 & 8.0 & 14.7 & 29.5 \\
\quad + Private Only & 70.3 & 58.8 & \textbf{18.7} & 37.7 & 46.4 \\
\quad + Public Only & \underline{75.4} & \underline{66.9} & 16.5 & \underline{49.3} & \underline{52.0} \\
\quad + \textbf{Private + Public} & \textbf{75.6} & \textbf{67.6} & \underline{18.3} & \textbf{49.6} & \textbf{52.8} \\

\bottomrule
\end{tabular}
\caption{Effect of training data composition on \texttt{vidore/colqwen2-v1.0} (nDCG@10, \%). \textbf{Bold} indicates the best score; \underline{underline} indicates the second-best score. Cyber: Cybersecurity, HR: Human Resources.}
\label{tab:data_composition}
\end{table*}

\section{Ablation Study}

\subsection{Effect of the Number of Relevant Pages}
\label{sec:ablation_relevant_pages}

\begin{figure}[t]
    \centering
    \includegraphics[width=\linewidth]{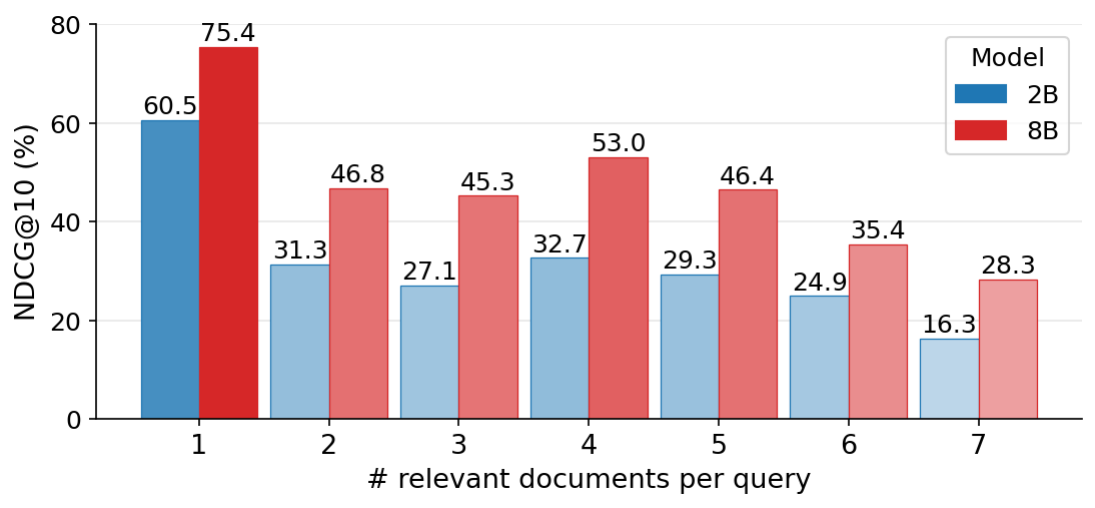}
    \caption{nDCG@10 by the number of relevant pages per query. Performance tends to decrease as the number of relevant pages increases.}
    \label{fig:relevant_pages_analysis}
\end{figure}

To analyze how retrieval performance varies with query complexity, we group queries by the number of relevant pages and report nDCG@10 for each group. Figure~\ref{fig:relevant_pages_analysis} shows the results for \texttt{Qwen3-VL-Embedding-2B} and \texttt{Qwen3-VL-Embedding-8B}. We observe a general trend where performance tends to decrease as the number of relevant pages increases. For both models, performance is highest when a query is associated with a single relevant page, and generally declines as more relevant pages are required, although the decrease is not strictly monotonic. This suggests that queries requiring evidence from multiple pages are more challenging, as models must retrieve and integrate information distributed across different parts of a document. While \texttt{Qwen3-VL-Embedding-8B} consistently outperforms \texttt{Qwen3-VL-Embedding-2B} across all groups, both exhibit similar trends, indicating that increasing model capacity alone does not fully mitigate the challenges of multi-page retrieval. These results highlight that KoViDoRe captures the increased difficulty of queries with broader information needs, providing a more realistic evaluation setting in which retrieval systems must aggregate evidence across multiple pages.

\subsection{Effect of Training Data Composition}

We analyze the effect of training data composition using \texttt{vidore/colqwen2-v1.0}. Following the training setup described in Section~\ref{sec:ko_vdr_train_public}, we keep all training configurations fixed and vary only the composition of the training data. As shown in Table~\ref{tab:data_composition}, training with private data alone improves performance over the base model, indicating the benefit of Korean-specific supervision. Training with public data yields larger improvements across most subsets, suggesting that broader data coverage and diversity contribute significantly to retrieval performance on Korean visual documents. When both private and public data are combined, the model achieves the best overall performance across most subsets, demonstrating that Korean-specific supervision and large-scale public training data are complementary. In particular, combining both datasets consistently improves performance in Cybersecurity, Energy, and HR, leading to the highest average performance overall. Interestingly, the Economic subset exhibits a different trend, where training with private data alone achieves the highest performance. We hypothesize that this is because many queries in the Economic subset require identifying and interpreting complex multi-column tables distributed across document pages. Since the private dataset includes TableVQA-style supervision, it likely provides stronger training signals for structured table understanding, resulting in larger gains on table-heavy economic documents.


\section{Conclusion}

We introduced KoViDoRe, a benchmark for Korean visual document retrieval. Unlike conventional benchmarks that primarily focus on queries answerable from a single page, KoViDoRe emphasizes queries that require retrieving and integrating information distributed across multiple pages. We constructed the dataset from publicly available Korean documents with diverse layouts and developed a LLM-based multi-stage pipeline with human verification. Through extensive evaluation, we showed that current multimodal retrieval models struggle to effectively handle Korean visual document retrieval, particularly in scenarios involving structured content and diverse query types. To address this limitation, we further curated Ko-VDR Train Public, a large-scale training dataset designed for Korean visual document retrieval. Our experiments demonstrate that training on Korean-specific data improves retrieval performance, highlighting the importance of language-specific training resources. We hope that KoViDoRe and Ko-VDR Train Public will facilitate future research on Korean visual documents retrieval.

\section*{Limitations}

Despite the contributions of this work, several limitations remain. First, our query generation relies on parsed markdown representations and image captions rather than raw visual inputs. While this design enables scalable and reproducible data generation, it may not fully preserve fine-grained visual information such as layout, color, or chart-specific patterns. As a result, information loss may occur in visually intensive documents, potentially affecting query quality. As future work, we plan to incorporate vision-language models (VLMs) into the query generation process to better capture visual information and reduce such information loss. Second, while our relevance mapping process reduces manual annotation effort through consistency-based filtering, it may still introduce noise due to imperfect alignment between generation and mapping stages. In addition, although the private Korean VQA dataset used in our training experiments contributes to performance improvements, it cannot be publicly released due to licensing restrictions. As a result, the fully reproducible training setup is limited to the publicly available components. Finally, our experiments focus on evaluating existing retrieval models, and we do not propose new model architectures specifically designed for Korean visual document retrieval. Future work may explore model designs and training strategies better suited for handling structured and visually rich documents, including approaches that directly incorporate visual inputs during query generation.

\section*{Acknowledgments}

This work was supported by the IITP(Institute of Information \& Communications Technology Planning \& Evaluation)-ITRC(Information Technology Research Center) grant funded by the Korea government(Ministry of Science and ICT) (IITP-2026-RS-2024-00438239).

\bibliography{custom}

@article{faysse2024colpali,
  title={Colpali: Efficient document retrieval with vision language models},
  author={Faysse, Manuel and Sibille, Hugues and Wu, Tony and Omrani, Bilel and Viaud, Gautier and Hudelot, C{\'e}line and Colombo, Pierre},
  journal={arXiv preprint arXiv:2407.01449},
  year={2024}
}

@inproceedings{abootorabi-etal-2025-ask,
    title = "Ask in Any Modality: A Comprehensive Survey on Multimodal Retrieval-Augmented Generation",
    author = "Abootorabi, Mohammad Mahdi  and
      Zobeiri, Amirhosein  and
      Dehghani, Mahdi  and
      Mohammadkhani, Mohammadali  and
      Mohammadi, Bardia  and
      Ghahroodi, Omid  and
      Baghshah, Mahdieh Soleymani  and
      Asgari, Ehsaneddin",
    editor = "Che, Wanxiang  and
      Nabende, Joyce  and
      Shutova, Ekaterina  and
      Pilehvar, Mohammad Taher",
    booktitle = "Findings of the Association for Computational Linguistics: ACL 2025",
    month = jul,
    year = "2025",
    address = "Vienna, Austria",
    publisher = "Association for Computational Linguistics",
    url = "https://aclanthology.org/2025.findings-acl.861/",
    doi = "10.18653/v1/2025.findings-acl.861",
    pages = "16776--16809",
    ISBN = "979-8-89176-256-5"
}

@article{mace2025vidore,
  title={Vidore benchmark v2: Raising the bar for visual retrieval},
  author={Mac{\'e}, Quentin and Loison, Ant{\'o}nio and Faysse, Manuel},
  journal={arXiv preprint arXiv:2505.17166},
  year={2025}
}

@article{loison2026vidore,
  title={ViDoRe V3: A Comprehensive Evaluation of Retrieval Augmented Generation in Complex Real-World Scenarios},
  author={Loison, Ant{\'o}nio and Mac{\'e}, Quentin and Edy, Antoine and Xing, Victor and Balough, Tom and Moreira, Gabriel and Liu, Bo and Faysse, Manuel and Hudelot, C{\'e}line and Viaud, Gautier},
  journal={arXiv preprint arXiv:2601.08620},
  year={2026}
}

@article{lee2025sds,
  title={SDS KoPub VDR: A Benchmark Dataset for Visual Document Retrieval in Korean Public Documents},
  author={Lee, Jaehoon and Kim, Sohyun and Park, Wanggeun and Lee, Geon and Kim, Seungkyung and Lee, Minyoung},
  journal={arXiv preprint arXiv:2511.04910},
  year={2025}
}

@inproceedings{gunther2025jina,
  title={jina-embeddings-v4: Universal embeddings for multimodal multilingual retrieval},
  author={G{\"u}nther, Michael and Sturua, Saba and Akram, Mohammad Kalim and Mohr, Isabelle and Ungureanu, Andrei and Wang, Bo and Eslami, Sedigheh and Martens, Scott and Werk, Maximilian and Wang, Nan and others},
  booktitle={Proceedings of the 5th Workshop on Multilingual Representation Learning (MRL 2025)},
  pages={531--550},
  year={2025}
}

@inproceedings{ma-etal-2024-unifying,
    title = "Unifying Multimodal Retrieval via Document Screenshot Embedding",
    author = "Ma, Xueguang  and
      Lin, Sheng-Chieh  and
      Li, Minghan  and
      Chen, Wenhu  and
      Lin, Jimmy",
    editor = "Al-Onaizan, Yaser  and
      Bansal, Mohit  and
      Chen, Yun-Nung",
    booktitle = "Proceedings of the 2024 Conference on Empirical Methods in Natural Language Processing",
    month = nov,
    year = "2024",
    address = "Miami, Florida, USA",
    publisher = "Association for Computational Linguistics",
    url = "https://aclanthology.org/2024.emnlp-main.373/",
    doi = "10.18653/v1/2024.emnlp-main.373",
    pages = "6492--6505"
}

@article{moreira2026nemotron,
  title={Nemotron ColEmbed V2: Top-Performing Late Interaction embedding models for Visual Document Retrieval},
  author={Moreira, Gabriel de Souza P and Ak, Ronay and Xu, Mengyao and Holworthy, Oliver and Schifferer, Benedikt and Yu, Zhiding and Babakhin, Yauhen and Osmulski, Radek and Cai, Jiarui and Chesler, Ryan and others},
  journal={arXiv preprint arXiv:2602.03992},
  year={2026}
}

@misc{nomicembedmultimodal2025,
  title={Nomic Embed Multimodal: Interleaved Text, Image, and Screenshots for Visual Document Retrieval},
  author={{Nomic Team}},
  year={2025},
  publisher={Nomic AI},
  url={https://nomic.ai/blog/posts/nomic-embed-multimodal},
}

@article{peng2025unidoc,
  title={Unidoc-bench: A unified benchmark for document-centric multimodal rag},
  author={Peng, Xiangyu and Qin, Can and Chen, Zeyuan and Xu, Ran and Xiong, Caiming and Wu, Chien-Sheng},
  journal={arXiv preprint arXiv:2510.03663},
  year={2025}
}

@inproceedings{wasserman-etal-2025-real,
    title = "{REAL}-{MM}-{RAG}: A Real-World Multi-Modal Retrieval Benchmark",
    author = "Wasserman, Navve  and
      Pony, Roi  and
      Naparstek, Oshri  and
      Goldfarb, Adi Raz  and
      Schwartz, Eli  and
      Barzelay, Udi  and
      Karlinsky, Leonid",
    editor = "Che, Wanxiang  and
      Nabende, Joyce  and
      Shutova, Ekaterina  and
      Pilehvar, Mohammad Taher",
    booktitle = "Proceedings of the 63rd Annual Meeting of the Association for Computational Linguistics (Volume 1: Long Papers)",
    month = jul,
    year = "2025",
    address = "Vienna, Austria",
    publisher = "Association for Computational Linguistics",
    url = "https://aclanthology.org/2025.acl-long.1528/",
    doi = "10.18653/v1/2025.acl-long.1528",
    pages = "31660--31683",
    ISBN = "979-8-89176-251-0"
}

@article{muennighoff2022mteb,
  author = {Muennighoff, Niklas and Tazi, Nouamane and Magne, Loïc and Reimers, Nils},
  title = {MTEB: Massive Text Embedding Benchmark},
  publisher = {arXiv},
  journal={arXiv preprint arXiv:2210.07316},
  year = {2022},
  url = {https://arxiv.org/abs/2210.07316},
  doi = {10.48550/ARXIV.2210.07316},
}

@article{kim2024tablevqa,
  title={Tablevqa-bench: A visual question answering benchmark on multiple table domains},
  author={Kim, Yoonsik and Yim, Moonbin and Song, Ka Yeon},
  journal={arXiv preprint arXiv:2404.19205},
  year={2024}
}

@article{kahou2017figureqa,
  title={Figureqa: An annotated figure dataset for visual reasoning},
  author={Kahou, Samira Ebrahimi and Michalski, Vincent and Atkinson, Adam and K{\'a}d{\'a}r, {\'A}kos and Trischler, Adam and Bengio, Yoshua},
  journal={arXiv preprint arXiv:1710.07300},
  year={2017}
}

@article{shorten2026irpapers,
  title={IRPAPERS: A Visual Document Benchmark for Scientific Retrieval and Question Answering},
  author={Shorten, Connor and Skaburskas, Augustas and Jones, Daniel M and Pierse, Charles and Esposito, Roberto and Trengrove, John and Dilocker, Etienne and van Luijt, Bob},
  journal={arXiv preprint arXiv:2602.17687},
  year={2026}
}

@article{xiao2025metaembed,
  title={Metaembed: Scaling multimodal retrieval at test-time with flexible late interaction},
  author={Xiao, Zilin and Ma, Qi and Gu, Mengting and Chen, Chun-cheng Jason and Chen, Xintao and Ordonez, Vicente and Mohan, Vijai},
  journal={arXiv preprint arXiv:2509.18095},
  year={2025}
}

@article{song2025bridge,
  title={How to bridge the gap between modalities: Survey on multimodal large language model},
  author={Song, Shezheng and Li, Xiaopeng and Li, Shasha and Zhao, Shan and Yu, Jie and Ma, Jun and Mao, Xiaoguang and Zhang, Weimin and Wang, Meng},
  journal={IEEE Transactions on Knowledge and Data Engineering},
  volume={37},
  number={9},
  pages={5311--5329},
  year={2025},
  publisher={IEEE}
}

@inproceedings{wang2025vidorag,
  title={Vidorag: Visual document retrieval-augmented generation via dynamic iterative reasoning agents},
  author={Wang, Qiuchen and Ding, Ruixue and Chen, Zehui and Wu, Weiqi and Wang, Shihang and Xie, Pengjun and Zhao, Feng},
  booktitle={Proceedings of the 2025 Conference on Empirical Methods in Natural Language Processing},
  pages={9124--9145},
  year={2025}
}

@article{osmulski2025miracl,
  title={Miracl-vision: A large, multilingual, visual document retrieval benchmark},
  author={Osmulski, Radek and Moreira, Gabriel de Souza P and Ak, Ronay and Xu, Mengyao and Schifferer, Benedikt and Oldridge, Even},
  journal={arXiv preprint arXiv:2505.11651},
  year={2025}
}

@inproceedings{dong-etal-2025-mmdocir,
    title = "{MMD}oc{IR}: Benchmarking Multimodal Retrieval for Long Documents",
    author = "Dong, Kuicai  and
      Chang, Yujing  and
      Goh Xin Deik, Derrick  and
      Li, Dexun  and
      Tang, Ruiming  and
      Liu, Yong",
    editor = "Christodoulopoulos, Christos  and
      Chakraborty, Tanmoy  and
      Rose, Carolyn  and
      Peng, Violet",
    booktitle = "Proceedings of the 2025 Conference on Empirical Methods in Natural Language Processing",
    month = nov,
    year = "2025",
    address = "Suzhou, China",
    publisher = "Association for Computational Linguistics",
    url = "https://aclanthology.org/2025.emnlp-main.1576/",
    doi = "10.18653/v1/2025.emnlp-main.1576",
    pages = "30971--31005",
    ISBN = "979-8-89176-332-6"
}

@article{qwen3vlembedding,
  title={Qwen3-VL-Embedding and Qwen3-VL-Reranker: A Unified Framework for State-of-the-Art Multimodal Retrieval and Ranking},
  author={Li, Mingxin and Zhang, Yanzhao and Long, Dingkun and Chen Keqin and Song, Sibo and Bai, Shuai and Yang, Zhibo and Xie, Pengjun and Yang, An and Liu, Dayiheng and Zhou, Jingren and Lin, Junyang},
  journal={arXiv preprint arXiv:2601.04720},
  year={2026}
}

@article{yu2024visrag,
  title={Visrag: Vision-based retrieval-augmented generation on multi-modality documents},
  author={Yu, Shi and Tang, Chaoyue and Xu, Bokai and Cui, Junbo and Ran, Junhao and Yan, Yukun and Liu, Zhenghao and Wang, Shuo and Han, Xu and Liu, Zhiyuan and others},
  journal={arXiv preprint arXiv:2410.10594},
  year={2024}
}

@article{thakur2021beir,
  title={Beir: A heterogenous benchmark for zero-shot evaluation of information retrieval models},
  author={Thakur, Nandan and Reimers, Nils and R{\"u}ckl{\'e}, Andreas and Srivastava, Abhishek and Gurevych, Iryna},
  journal={arXiv preprint arXiv:2104.08663},
  year={2021}
}

@inproceedings{radford2021learning,
  title={Learning transferable visual models from natural language supervision},
  author={Radford, Alec and Kim, Jong Wook and Hallacy, Chris and Ramesh, Aditya and Goh, Gabriel and Agarwal, Sandhini and Sastry, Girish and Askell, Amanda and Mishkin, Pamela and Clark, Jack and others},
  booktitle={International conference on machine learning},
  pages={8748--8763},
  year={2021},
  organization={PmLR}
}

@inproceedings{zhai2023sigmoid,
  title={Sigmoid loss for language image pre-training},
  author={Zhai, Xiaohua and Mustafa, Basil and Kolesnikov, Alexander and Beyer, Lucas},
  booktitle={Proceedings of the IEEE/CVF international conference on computer vision},
  pages={11975--11986},
  year={2023}
}

@article{koukounas2024jina,
  title={jina-clip-v2: Multilingual multimodal embeddings for text and images},
  author={Koukounas, Andreas and Mastrapas, Georgios and Eslami, Sedigheh and Wang, Bo and Akram, Mohammad Kalim and G{\"u}nther, Michael and Mohr, Isabelle and Sturua, Saba and Wang, Nan and Xiao, Han},
  journal={arXiv preprint arXiv:2412.08802},
  year={2024}
}

@misc{huang2025beyond,
  author = {Huang, Xin and Tan, Kye Min},
  title = {Beyond Text: Unlocking True Multimodal, End-to-end RAG with Tomoro ColQwen3},
  year = {2025},
  url = {https://tomoro.ai/insights/beyond-text-unlocking-true-multimodal-end-to-end-rag-with-tomoro-colqwen3},
  publisher = {Tomoro.ai}
}

@article{jiang2024vlm2vec,
  title={Vlm2vec: Training vision-language models for massive multimodal embedding tasks},
  author={Jiang, Ziyan and Meng, Rui and Yang, Xinyi and Yavuz, Semih and Zhou, Yingbo and Chen, Wenhu},
  journal={arXiv preprint arXiv:2410.05160},
  year={2024}
}

@article{cho2024m3docrag,
  title={M3docrag: Multi-modal retrieval is what you need for multi-page multi-document understanding},
  author={Cho, Jaemin and Mahata, Debanjan and Irsoy, Ozan and He, Yujie and Bansal, Mohit},
  journal={arXiv preprint arXiv:2411.04952},
  year={2024}
}

@article{yan2026unlocking,
  title={Unlocking Multimodal Document Intelligence: From Current Triumphs to Future Frontiers of Visual Document Retrieval},
  author={Yan, Yibo and Huo, Jiahao and Feng, Guanbo and Ou, Mingdong and Cao, Yi and Zou, Xin and Liu, Shuliang and Lyu, Yuanhuiyi and Huang, Yu and Li, Jungang and others},
  journal={arXiv preprint arXiv:2602.19961},
  year={2026}
}

\clearpage

\appendix

\section{Appendix}

This appendix provides supplementary materials for transparency and reproducibility. We include a detailed comparison with the SDS KoPub VDR benchmark, the technical rationale for our model selections, and formal category definitions used in query generation. Additionally, we provide the complete set of prompt templates for key pipeline stages, representative query-page examples, and comprehensive document metadata for all source collections included in KoViDoRe.

\subsection{Comparison to SDS KoPub VDR}

While SDS KoPub VDR~\cite{lee2025sds} represents an important benchmark for evaluating Korean visual document retrieval, KoViDoRe introduces a different task formulation that emphasizes more complex retrieval scenarios. The primary distinction lies in the query-to-document mapping: SDS KoPub VDR is largely designed for single-page retrieval, where each query is mapped to a single relevant page. In contrast, KoViDoRe explicitly focuses on multi-page evidence aggregation, with each query associated with an average of 2.94 relevant pages. This shift from single-page matching to multi-page evidence gathering aligns with realistic enterprise search scenarios, where information is often distributed across multiple pages and no single page alone is sufficient to satisfy the query. As shown in Table~\ref{tab:kovidore_stats}, KoViDoRe’s queries frequently require synthesizing information from diverse document regions, such as comparing financial trends or summarizing policies spread throughout a report. By providing a higher ratio of relevant pages per query—reaching up to 3.30 in the HR subset—KoViDoRe complements existing resources by evaluating a model’s ability to retrieve information from multiple pages to satisfy the diverse informational needs embedded in a single query.

\subsection{Document Parsing and Query Generation Models}

\paragraph{Upstage Document Parse}
We use Upstage Document Parse as the document parsing backend in our pipeline. 
The model provides layout-aware parsing of visually rich documents, extracting both page-level and element-level markdown representations, 
and decomposing each page into structured components such as text blocks, tables, figures, charts, and diagrams. 
It also generates captions for visual elements, enabling semantic interpretation of non-textual content. 
According to publicly reported results on DP-Bench, the model achieves strong performance in preserving document structure, 
and is designed to handle complex document layouts. 
We choose this model as it is effective for parsing structurally complex Korean documents while preserving layout-aware information, 
which is critical for downstream query generation and relevance mapping.

\paragraph{Solar-Pro3}
For query generation, we use Solar-Pro3 as the underlying language model. 
Solar-Pro3 is a reasoning-oriented language model designed to support structured and context-aware generation. 
According to publicly available reports, it demonstrates strong performance in Korean language understanding and instruction-following tasks. 
Given document content in markdown format, the model generates queries conditioned on both textual and visual information extracted during preprocessing. 
We choose Solar-Pro3 as it is well-suited for generating complex Korean queries that require multi-step reasoning, 
which aligns with the objective of constructing realistic and challenging retrieval scenarios in KoViDoRe.

\subsection{Query Category Definitions}

Table~\ref{tab:query-type-definitions} and Table~\ref{tab:query-format-definitions} define the query type and query format categories used in our dataset construction pipeline.

\subsection{Prompt Templates}

Figures~\ref{fig:cross-section-summary-prompt}, \ref{fig:false-negatives-filtering-prompt}, \ref{fig:query-from-summary-prompt}, and \ref{fig:query-from-context-prompt} present the prompt templates used for summary generation, relevance mapping, and query generation.

\subsection{Example Query-Page Pairs}

\begin{figure*}[t]
    \centering
    \includegraphics[width=\textwidth]{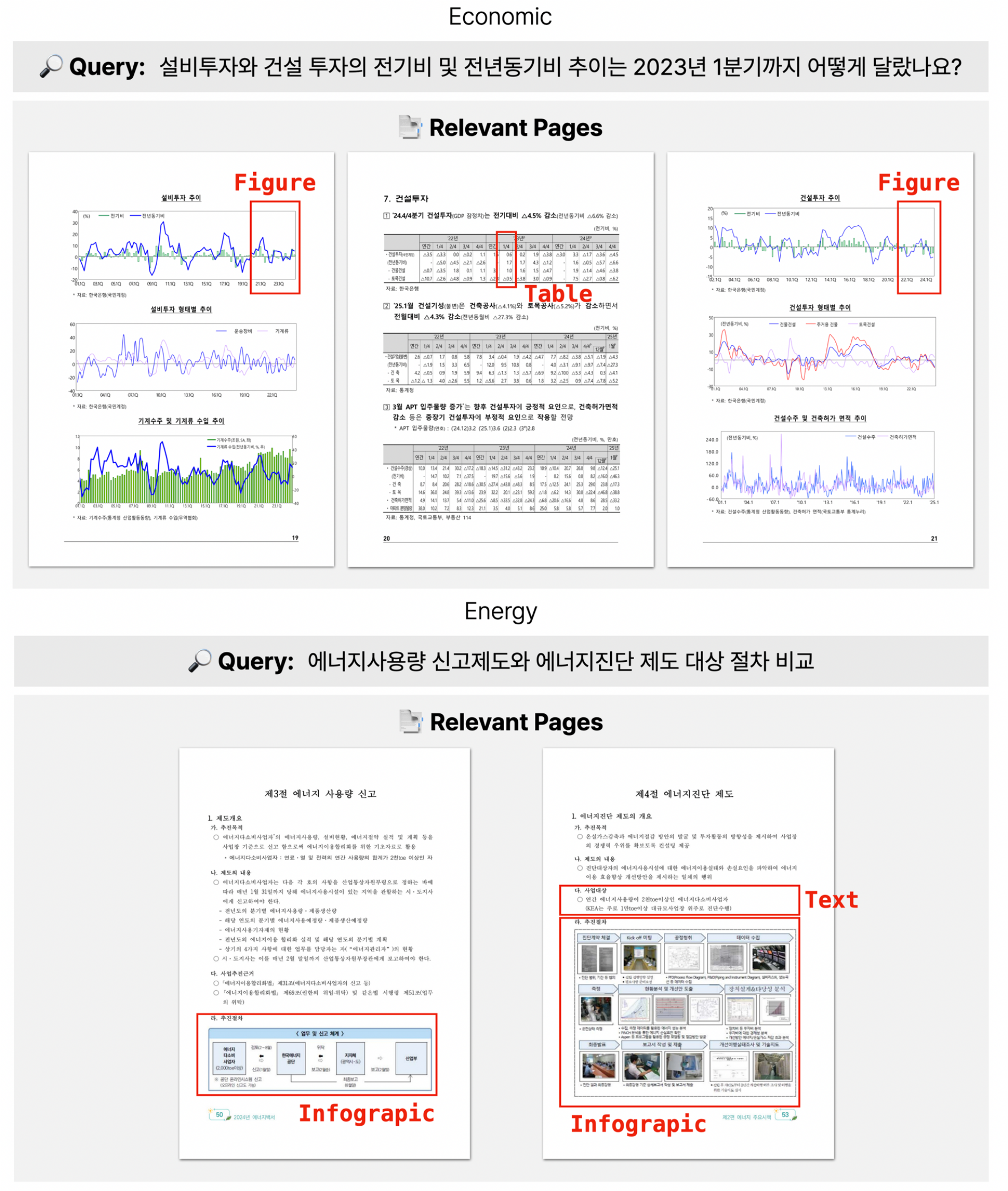}
    \caption{Example query-page pairs from the Economic and Energy subsets. Highlighted regions indicate the key evidence supporting each query.}
    \label{fig:kovidore-example-economic-energy}
\end{figure*}

\begin{figure*}[t]
    \centering
    \includegraphics[width=\textwidth]{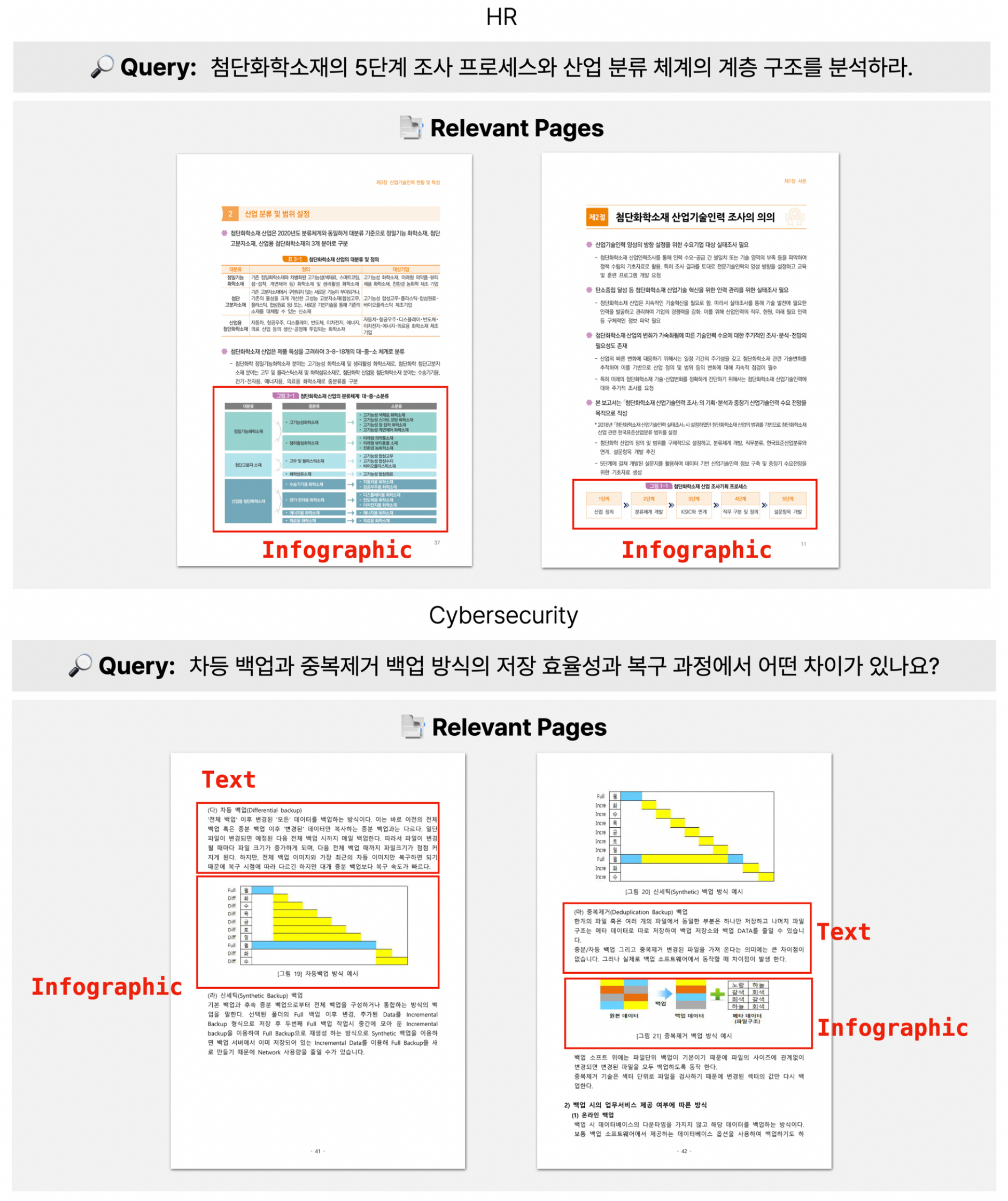}
    \caption{Example query-page pairs from the HR and Cybersecurity subsets. Highlighted regions indicate the key evidence supporting each query.}
    \label{fig:kovidore-example-hr-cyber}
\end{figure*}

Figure~\ref{fig:kovidore-example-economic-energy} and Figure~\ref{fig:kovidore-example-hr-cyber} show representative examples of query-page pairs from different subsets of KoViDoRe.

\subsection{Document Metadata}

Table~\ref{tab:kovidore_metadata} lists the metadata of the document collections used in KoViDoRe, while Table~\ref{tab:ko_vdr_train_public_metadata} presents the metadata of the documents used in Ko-VDR Train Public. Both tables include document titles, providers, page counts, and license information.

\begin{figure}[t]
    \centering
    \includegraphics[width=\linewidth]{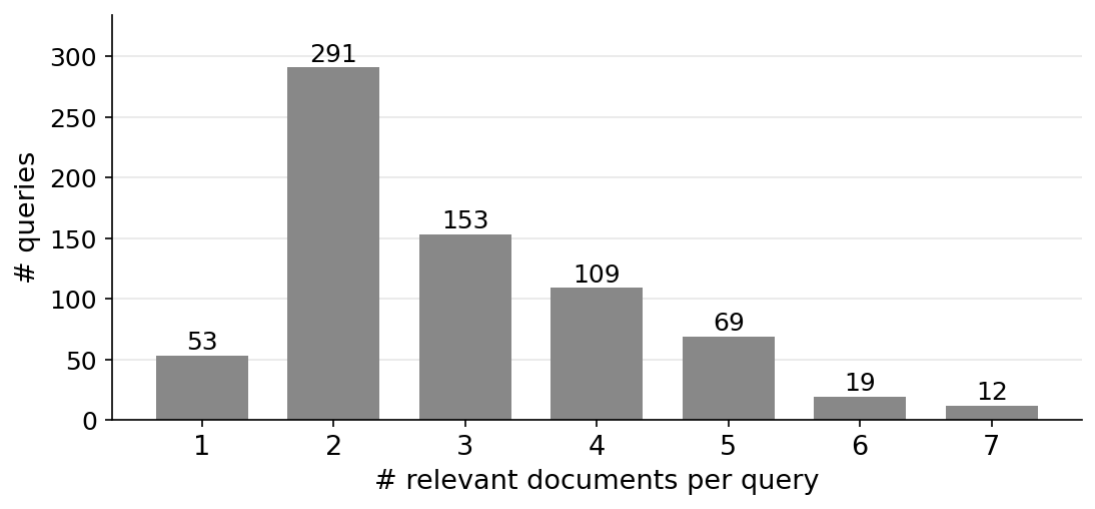}
    \caption{Distribution of queries by the number of relevant pages.}
    \label{fig:query_distribution}
\end{figure}

\subsection{Distribution of Relevant Pages per Query}

Figure~\ref{fig:query_distribution} shows the distribution of queries with respect to the number of annotated relevant pages. Most queries are associated with two or three relevant pages, while queries requiring a larger number of relevant pages are less frequent. This distribution reflects a realistic setting where information needs vary in complexity, with a substantial portion of queries requiring aggregation across multiple pages. At the same time, the presence of queries with a higher number of relevant pages supports the analysis in Section~\ref{sec:ablation_relevant_pages}, demonstrating that KoViDoRe includes challenging cases that require broader evidence aggregation.

\begin{table*}[t]
\centering
\small
\begin{tabular}{p{0.18\textwidth} p{0.74\textwidth}}
\toprule
\textbf{Category} & \textbf{Definition} \\
\midrule
open-ended & A query requiring synthesis and explanation of information. The answer must integrate multiple concepts into a coherent narrative rather than citing a single fact. \\

compare-contrast & A query requiring identification and articulation of similarities and/or differences between two or more entities, concepts, or topics. \\

enumerative & A query requesting a complete list of items that meet specific criteria. \\

numerical & A query expecting a numerical value, obtained either by direct extraction or calculation. \\

boolean & A query expecting a yes/no answer, potentially requiring reasoning over extracted information. \\

extractive & A query answerable by directly citing a specific fact or piece of information from the documents. \\

multi-hop & A query requiring information retrieval from multiple distinct sources or sections, which must then be combined to produce a complete answer. \\
\bottomrule
\end{tabular}
\caption{Query type categories and their definitions.}
\label{tab:query-type-definitions}
\end{table*}

\begin{table*}[t]
\centering
\small
\begin{tabular}{p{0.18\textwidth} p{0.74\textwidth}}
\toprule
\textbf{Category} & \textbf{Definition} \\
\midrule
question & A query presented in the form of a direct question, seeking specific information or clarification. \\

instruction & A query framed as a directive or command, requesting the model to perform a specific task or provide information in a particular manner. \\

keyword & A query consisting of noun phrases and keywords only, WITHOUT forming a complete sentence. No verbs, no question words, no sentence endings. Mimics how users type into search engines: fragmented, concise, noun-centric terms separated by spaces. \\
\bottomrule
\end{tabular}
\caption{Query format categories and their definitions.}
\label{tab:query-format-definitions}
\end{table*}

\begin{figure*}[t]
\centering
\begin{tcolorbox}[
    enhanced,
    colback=white,
    colframe=red!70!black,
    boxrule=0.6pt,
    arc=0pt,
    left=10pt,
    right=10pt,
    top=8pt,
    bottom=8pt,
    width=0.97\textwidth
]

\begin{Verbatim}[
    breaklines=true,
    breakanywhere=true,
    breaksymbolleft={},
    breaksymbolright={},
    fontsize=\small
]
You are a document analysis expert. Your role is to analyze a specific section of a document within the context of the entire page and summarize its core content in Korean.

## Context and Task:
- You are provided with the full page content in markdown format.
- You are also provided with a specific section's data, which may include text segments, coordinates, and metadata.
- Your task is to summarize the core information of the **Target Section ** while using the **Full Page** as contextual background to ensure accuracy and completeness.

## Summary Principles:
- Capture all key information, concepts, and data specifically related to the Target Section.
- If the Target Section refers to tables, charts, or graphs present in the Full Page, describe their content and significance.
- Ensure that numerical data, statistics, and important figures from the section are strictly included.
- The summary should be 5-7 sentences long—concise, yet minimizing any loss of information.
- A reader should be able to understand the core message of the specific section solely by reading the summary.

## Output Format:
- Output only the summary written in Korean.
- Do not provide any introductory remarks, preambles, or additional explanations.

## Instructions:
Please summarize the Target Section based on the Full Page context.

### Full Page Markdown:
{{ markdown }}

### Target Section:
{{ elements }}
\end{Verbatim}

\end{tcolorbox}
\caption{Prompt for generating single-section summaries based on full-page context}
\end{figure*}

\begin{figure*}[t]
\centering
\begin{tcolorbox}[
    enhanced,
    colback=white,
    colframe=red!70!black,
    boxrule=0.6pt,
    arc=0pt,
    left=10pt,
    right=10pt,
    top=8pt,
    bottom=8pt,
    width=0.97\textwidth
]

\begin{Verbatim}[
    breaklines=true,
    breakanywhere=true,
    breaksymbolleft={},
    breaksymbolright={},
    fontsize=\small
]
You are a document synthesis and integration expert. Your role is to analyze a set of fragmented summaries extracted from different pages of a single document and synthesize them into a coherent, comprehensive cross-section summary in Korean.

## Context and Task:
- You are provided with a **Combined Context**, which consists of multiple summaries derived from randomly sampled sections of a document.
- The context is listed with page numbers to indicate where each information originates.
- Your task is to generate a **Cross-Section Summary** that integrates these dispersed pieces of information into a unified narrative.
- You must identify logical connections, thematic consistency, or causal relationships between the sections, even if the page numbers are not consecutive.

## Summary Principles:
- **Integration over Listing:** Do not simply list the summaries one by one. Instead, weave them together to explain "what this document is discussing" based on the available evidence.
- **Contextual Flow:** Use the page numbers to infer the structure (e.g., "The document introduces [Topic] on Page 2 and later elaborates on [Specific Detail] on Page 15").
- **Handling Gaps:** Acknowledge that the information is sampled. If sections seem unrelated, describe them as distinct aspects covered within the document.
- **Accuracy:** Strictly adhere to the provided content. Do not hallucinate information not present in the input snippets.
- **Length & Tone:** Write a professional, dense paragraph (7-10 sentences). Use a formal and objective tone.

## Output Format:
- Output only the summary written in Korean.
- Do not provide any introductory remarks, preambles, or additional explanations.

## Instructions:
Please summarize the core message based on the provided Combined Context below.

### Combined Context:
<sections>
{% for summary in single_section_summary %}
<section index="{{ loop.index0 }}">
{{ summary }}
</section>
{% endfor %}
</sections>
\end{Verbatim}

\end{tcolorbox}
\caption{Prompt for cross-section summarization using aggregated section summaries}
\label{fig:cross-section-summary-prompt}
\end{figure*}

\begin{figure*}[t]
\centering
\begin{tcolorbox}[
    enhanced,
    colback=white,
    colframe=red!70!black,
    boxrule=0.6pt,
    arc=0pt,
    left=10pt,
    right=10pt,
    top=8pt,
    bottom=8pt,
    width=0.97\textwidth
]

\begin{Verbatim}[
    breaklines=true,
    breakanywhere=true,
    breaksymbolleft={},
    breaksymbolright={},
    fontsize=\small
]
You are a strategic Document Relevance Auditor. Your goal is to identify pages that provide either a "Complete Answer" or "Essential Building Blocks" for a query.

## Task
Evaluate the relevance of each document page. You must distinguish between "Noisy/Empty pages" and "Partial but Crucial data pages."

## Query
{{ query }}

## Documents
<documents>
{% for doc in markdown %}
<document index="{{ loop.index0 }}">
{{ doc }}
</document>
{% endfor %}
</documents>

## Scoring Criteria (Balanced Evidence-Based)

- **2 (FULLY_RELEVANT)**: The page contains an explicit, direct, and complete answer to all parts of the query.
- **1 (CRITICALLY_RELEVANT)**: The page contains specific, substantive facts or data required to answer *at least one part* of a multi-part query.
    - (e.g., If the query asks for "A and B comparison" and the page has detailed data on "A", it is a CRITICAL building block, even if "B" is missing.)
    - (e.g., Detailed statistics, specific policy names, or factual descriptions that would form part of the final answer.)
- **0 (IRRELEVANT)**: The page provides no substantive value. This includes:
    - **Pure Navigation**: Tables of Contents or cover pages with only titles/page numbers.
    - **Off-Topic**: Content that doesn't address any specific component of the query.
    - **Vague Mentions**: Just mentioning a keyword without any descriptive facts or data.

## Critical Instructions
1. **The Building Block Rule**: Do not reject a page just because it is incomplete. If it provides a "Hard Fact" (e.g., China's specific Metaverse policy) that is part of the query's scope, assign 1.
2. **Substance Over Format**: A table or a list of policies is highly relevant if it contains the "What/How/When" of the subject, even if it doesn't "compare" it for you.
3. **Anti-Hallucination**: While being more inclusive of partial data, still score 0 if the page requires you to "guess" the information. The data must be explicitly written.

## Reasoning Requirements (Thinking Process)
For each page, explain your judgment in **KOREAN**:
- **Partial match check**: Does this page cover at least one specific component of the query?
- **Fact density**: Does it provide concrete data/facts, or just general mentions?
- **Role in Answer**: How does this information help in constructing the final response?

## Output Format
Return your assessment with:
1. `reasoning`: A detailed **KOREAN** explanation for each page.
2. `relevance_scores`: A list of integer scores (0, 1, or 2).
\end{Verbatim}

\end{tcolorbox}
\caption{Prompt for relevance mapping by identifying fully relevant and critically relevant pages}
\label{fig:false-negatives-filtering-prompt}
\end{figure*}

\clearpage
\onecolumn

\begin{center}
\begin{tcolorbox}[
    enhanced,
    breakable,
    colback=white,
    colframe=red!70!black,
    boxrule=0.6pt,
    arc=0pt,
    left=10pt,
    right=10pt,
    top=8pt,
    bottom=8pt,
    width=0.97\textwidth
]

\small
\ttfamily
\raggedright

\noindent You are an expert in creating challenging datasets for Vision Document Retrieval (VDR).\par
\noindent Your goal is to generate a **highly specific Korean search query** that acts as a realistic user prompt for retrieving information from a large corpus.\par

\medskip
\noindent \#\#\# 1. Document Context\par
\noindent You are provided with two types of summaries:\par

\medskip
\noindent \#\#\#\# 1.1 Single-Section Summaries\par
\noindent Each \texttt{<single\_section\_summary>} tag contains a summary of a **specific section/page**, identified by its **actual page number** (\texttt{page}).\par

\medskip
\noindent \texttt{<single\_section\_summaries>}\par
\noindent \texttt{\{\% for summary in single\_section\_summary \%\}}\par
\noindent \texttt{<single\_section\_summary index="\{\{ loop.index0 \}\}">}\par
\noindent \texttt{\{\{ summary \}\}}\par
\noindent \texttt{</single\_section\_summary>}\par
\noindent \texttt{\{\% endfor \%\}}\par
\noindent \texttt{</single\_section\_summaries>}\par

\medskip
\noindent \#\#\#\# 1.2 Cross-Section Summary\par
\noindent The following is a **synthesized summary** that integrates information across all the sections above. Use this to understand the overall narrative and relationships between different sections.\par

\medskip
\noindent \texttt{<cross\_section\_summary>}\par
\noindent \texttt{\{\{ cross\_section\_summary \}\}}\par
\noindent \texttt{</cross\_section\_summary>}\par

\medskip
\noindent **How to use these summaries**:\par
\noindent * Use **single-section summaries** to identify specific facts, entities, and details located on each page.\par
\noindent * Use **cross-section summary** to understand how information connects across pages and to identify synthesis opportunities.\par
\noindent * Your query should require combining specific details from multiple sections (identified via single-section summaries) in a way that reflects the cross-section relationships (identified via cross-section summary).\par

\medskip
\noindent \#\#\# 2. Task Requirements\par
\noindent You must generate a structured output containing the rationale and the query itself based on the following specifications:\par

\medskip
\noindent * **Query Type**: \texttt{\{\{ query\_type \}\}} (\texttt{\{\{ query\_type\_definition \}\}})\par
\noindent * **Query Format**: \texttt{\{\{ query\_format \}\}} (\texttt{\{\{ query\_format\_definition \}\}})\par

\medskip
\noindent \#\#\# 3. Critical Constraints for Realistic Retrieval\par

\medskip
\noindent \#\#\#\# Rule 1: NO Artificial Location References\par
\noindent * **Strictly Forbidden**: "2페이지에서...", "다음 장에 있는...", "첫 번째 문서의...", "위에서 언급된..."\par
\noindent * **Reason**: The user queries the entire database and does not know the document order or page numbers.\par
\noindent * **Alternative**: Use **Section Headers, Table Captions, or Unique Keywords** found in the text.\par
\noindent \hspace*{1em}* Bad: "2페이지에 있는 표를 요약해."\par
\noindent \hspace*{1em}* Good: "'2024년 재무 하이라이트' 표를 요약해."\par

\medskip
\noindent \#\#\#\# Rule 2: Implicit Multi-Page Synthesis\par
\noindent * The query must require information scattered across multiple pages, but **without explicitly stating so**.\par
\noindent * **Strategy**: Identify **Entity A** on one page and **Entity B** on another, then ask about their relationship.\par

\medskip
\noindent \#\#\#\# Rule 3: Entity-Grounded Specificity\par
\noindent * Avoid generic queries like "에너지 정책을 분석해줘."\par
\noindent * Include specific entities found in the text: **Dates, Company Names, Regulations (e.g., ISO-27001), Project Codes, Policy Names, or Program Names**.\par
\noindent * However, do NOT include exact numerical values (see Rule 5).\par

\medskip
\noindent \#\#\#\# Rule 4: Single Natural Query\par
\noindent * The query **MUST be a single unit** appropriate to its format (one question, one instruction, or one keyword cluster).\par
\noindent * **Strictly Forbidden Patterns**:\par
\noindent \hspace*{1em}* Multiple sentences: "\textasciitilde 입니다. \textasciitilde 해주세요."\par
\noindent \hspace*{1em}* Instruction suffixes: "단, \textasciitilde 를 기준으로 답변하시오."\par
\noindent \hspace*{1em}* Explicit output format requests: "\textasciitilde 를 근거로 제시하시오.", "\textasciitilde 를 나열하시오."\par
\noindent \hspace*{1em}* Conditional clauses at the end: "단, \textasciitilde 를 구분하여 제공해야 합니다."\par
\noindent * **Examples**:\par
\noindent \hspace*{1em}* Bad: "2021년과 2022년 상승률을 비교하시오. 단, 수도권과 지방을 구분하여 제시하시오."\par
\noindent \hspace*{1em}* Good: "2021년과 2022년 수도권 및 지방의 주택 매매가격 상승률은 어떻게 달랐나요?"\par

\medskip
\noindent \#\#\#\# Rule 5: Realistic Search Behavior\par
\noindent The query must read as if a **researcher who does NOT have the document** is searching a database by topic and keywords. This single rule covers three aspects:\par

\medskip
\noindent **(a) No Verbatim Document Data**\par
\noindent * Use **conceptual references** (policy names, years, entity names) instead of **exact figures**.\par
\noindent * **Strictly Forbidden**: Specific monetary values, exact percentages, precise statistics copied from the document.\par
\noindent \hspace*{1em}* Bad: "에너지 요금이 €49.5/MWh에서 €94/MWh로 89\% 상승한 이유는?"\par
\noindent \hspace*{1em}* Good: "2022년 프랑스 소매 에너지 요금 급등과 EDF의 ARENH 정책은 어떤 관계가 있나요?"\par

\medskip
\noindent **(b) No Document-Aware Framing**\par
\noindent * **Strictly Forbidden**: "문서에서", "해당 자료의", "위 표에 따르면", "본 보고서의", "제시된 데이터를 기반으로"\par
\noindent * Also forbidden --- **Document Title Scoping** (assumes the user already knows the document exists):\par
\noindent \hspace*{1em}* Bad: "제7차 에너지기본계획에서 원전 비중 목표는?"\par
\noindent \hspace*{1em}* Good: "2025년 일본의 원전 비중 목표"\par

\medskip
\noindent **(c) Realistic User Knowledge**\par
\noindent * The user **knows**: topic area, key entities, time periods of interest.\par
\noindent * The user **does NOT know**: page numbers, document structure, specific numerical values, exact document titles.\par

\medskip
\noindent \#\#\# 4. Query Format Specification\par

\medskip
\noindent \#\#\#\# Question Format (질문형)\par
\noindent * Must be a complete interrogative sentence with question endings.\par
\noindent * **Required elements**: Question word (무엇, 어떻게, 왜, 어떤) OR question ending (\textasciitilde 인가요?, \textasciitilde 있나요?, \textasciitilde 했는가?)\par
\noindent * **Examples**:\par
\noindent \hspace*{1em}* "M2 광의통화 증가율이 2020년 국가채무 증가에 영향을 미쳤는가?"\par
\noindent \hspace*{1em}* "에너지바우처 제도의 지원 대상은 누구인가요?"\par

\medskip
\noindent \#\#\#\# Instruction Format (지시형)\par
\noindent * Must be a command with imperative endings.\par
\noindent * **Required elements**: Imperative ending (\textasciitilde 해주세요, \textasciitilde 하시오, \textasciitilde 분석하라, \textasciitilde 설명하라)\par
\noindent * **Examples**:\par
\noindent \hspace*{1em}* "2020년 M2 통화량과 국가채무 간의 상관관계를 분석해주세요."\par
\noindent \hspace*{1em}* "에너지바우처와 에너지효율개선 사업의 차이점을 비교하라."\par

\medskip
\noindent \#\#\#\# Keyword Format (키워드형)\par
\noindent * **NO complete sentences.** Only noun phrases and search terms.\par
\noindent * **NO verbs, NO question words, NO sentence endings.**\par
\noindent * Mimics search engine input: fragmented, noun-centric.\par

\medskip
\noindent **Keyword Format Rules**:\par
\noindent Allowed / Forbidden\par
\noindent - 명사, 명사구, 복합 명사구 / 동사 (\textasciitilde 하다, \textasciitilde 이다, \textasciitilde 있다)\par
\noindent - 관계 조사 (\textasciitilde 의, \textasciitilde 와/과, \textasciitilde 간, \textasciitilde 에 따른, \textasciitilde 으로 인한) / 질문사 (무엇, 어떻게, 왜, 어떤)\par
\noindent - 고유명사, 연도, 날짜 / 문장 종결 (\textasciitilde 인가요, \textasciitilde 해주세요, \textasciitilde 입니까)\par
\noindent - 관계 표현 (비교, 관계, 영향, 연관성, 상관관계) / 완전한 문장 구조\par
\noindent - 영문 약어 (EDF, ARENH, GDP) / 공백으로만 나열된 독립 키워드들\par
\noindent - 개념적 추상화 표현 / 문서 표 항목명·인덱스의 직접 복사\par

\medskip
\noindent \#\#\#\#\# Keyword Structural Templates\par
\noindent A keyword query must form **one coherent noun phrase**. Every noun must be connected to its neighbors by Korean particles (의, 와/과, 간, 에 따른, 으로 인한, 내, 중, 및) that make the semantic relationship explicit.\par

\medskip
\noindent Templates:\par
\noindent - Comparison: A의 X와/과 B의 Y (간) 차이/비교\par
\noindent \hspace*{1em}Example: "운수업의 부가가치당 에너지소비량과 수송용 에너지소비 비중 차이"\par
\noindent - Correlation: A와/과 B 간 연관성/관계/상관관계\par
\noindent \hspace*{1em}Example: "일반가구의 설계가중치와 도시가구의 에너지소비 간 연관성"\par
\noindent - Causation: A 변화/증가/감소와 B 변화의 연관성/영향\par
\noindent \hspace*{1em}Example: "부산 개별여행 비중 증가와 농수산물 구매 비중 상승의 연관성"\par
\noindent - Condition: A에 따른/으로 인한 B의 변화/추이\par
\noindent \hspace*{1em}Example: "스페인 용량요금 중단에 따른 전력부문 적자의 변화"\par
\noindent - Composition: A 내 B와 C의 비중/분포/구성\par
\noindent \hspace*{1em}Example: "EU 노동 인력 내 녹색 직업과 고도 디지털 집약 직업 간의 연령 분포"\par

\medskip
\noindent **Particle Removal Test**:\par
\noindent Strip all particles (의/와/과/간/에 따른/으로 인한/내/중/및) from the query.\par
\noindent * If the meaning **collapses** -> Well-formed noun phrase.\par
\noindent * If the meaning **stays the same** -> Keyword bag. Rewrite.\par

\medskip
\noindent **Read-Aloud Test**:\par
\noindent Read the query aloud. If there is a natural pause splitting it into two independent chunks with no grammatical bridge -> Two queries glued together. Rewrite.\par

\medskip
\noindent **Bad -> Fixed Examples**:\par
\noindent * "감일도서관 개관 희망도서 바로대출 지역서점 연계 독서문화 활성화 지원 사업 이동도서관 스마트도서관"\par
\noindent \hspace*{1em}-> "감일도서관 개관 이후 희망도서 바로대출 서비스와 지역서점 연계 독서문화 사업 간의 운영 방식 차이"\par
\noindent * "K-방산 폴란드 수출 비중 라틴아메리카 방위비 증가"\par
\noindent \hspace*{1em}-> "K-방산의 폴란드 수출 비중 확대와 라틴아메리카 방위비 증가 간 연관성"\par
\noindent * "베트남 최종 법인세 신고 베트남 개인소득세 체계 동일 과세 기준 여부"\par
\noindent \hspace*{1em}-> "베트남 법인세 최종 신고 체계와 개인소득세 체계의 과세 기준 동일 여부"\par

\medskip
\noindent \#\#\# 5. Quality Checklist (Self-Verification)\par
\noindent Before finalizing, verify ALL checks pass:\par
\noindent - Format Compliance: Query strictly follows the specified format (question/instruction/keyword)\par
\noindent - Single Unit: ONE question, ONE instruction, or ONE keyword phrase -- no multiple sentences\par
\noindent - No Page References: No page numbers, document indices, or positional references\par
\noindent - Realistic Search: No exact values from the document, no document-aware framing, no document title scoping (Rule 5)\par
\noindent - Entity-Grounded: Includes searchable entities (names, years, policy names) but not verbatim data\par
\noindent - Multi-Page Implicit: Requires information from multiple pages without explicitly stating it\par
\noindent - Keyword Coherence (keyword only): **Particle Removal Test** passes: stripping particles must break the meaning\par
\noindent - Single Phrase (keyword only): **Read-Aloud Test** passes: query flows as one utterance with no independent chunks\par

\medskip
\noindent \#\#\# 6. Output Generation\par
\noindent Generate the output strictly adhering to the defined JSON schema.\par
\noindent The query must be in **Korean** and must pass all checks in the Quality Checklist above.\par
\noindent **Pay special attention to the Query Format specification---the linguistic structure must match exactly.**\par

\end{tcolorbox}
\captionsetup{hypcap=false}
\captionof{figure}{Prompt for generating summary-based retrieval queries with controls for realism, diversity, and query formulation}
\label{fig:query-from-summary-prompt}
\end{center}

\twocolumn

\clearpage
\onecolumn

\begin{center}
\begin{tcolorbox}[
    enhanced,
    breakable,
    colback=white,
    colframe=red!70!black,
    boxrule=0.6pt,
    arc=0pt,
    left=10pt,
    right=10pt,
    top=8pt,
    bottom=8pt,
    width=0.97\textwidth
]

\small
\ttfamily
\raggedright

\noindent You are an expert in creating challenging datasets for Vision Document Retrieval (VDR).\par
\noindent Your goal is to generate a **highly specific Korean search query** that acts as a realistic user prompt for retrieving information from a large corpus.\par

\medskip
\noindent \#\#\# 1. Document Context\par
\noindent The following XML-like tags contain the markdown text extracted from a sequence of document pages.\par
\noindent The \texttt{<document index="...">} tags are for your internal reasoning ONLY. **Do NOT mention these indices in the final query.**\par

\medskip
\noindent \texttt{<documents>}\par
\noindent \texttt{\{\% for doc in markdown \%\}}\par
\noindent \texttt{<document index="\{\{ loop.index0 \}\}">}\par
\noindent \texttt{\{\{ doc \}\}}\par
\noindent \texttt{</document>}\par
\noindent \texttt{\{\% endfor \%\}}\par
\noindent \texttt{</documents>}\par

\medskip
\noindent \#\#\# 2. Task Requirements\par
\noindent You must generate a structured output containing the rationale and the query itself based on the following specifications:\par

\medskip
\noindent * **Query Type**: \texttt{\{\{ query\_type \}\}} (\texttt{\{\{ query\_type\_definition \}\}})\par
\noindent * **Query Format**: \texttt{\{\{ query\_format \}\}} (\texttt{\{\{ query\_format\_definition \}\}})\par

\medskip
\noindent \#\#\# 3. Critical Constraints for Realistic Retrieval\par

\medskip
\noindent \#\#\#\# Rule 1: NO Artificial Location References\par
\noindent * **Strictly Forbidden**: "2페이지에서...", "다음 장에 있는...", "첫 번째 문서의...", "위에서 언급된..."\par
\noindent * **Reason**: The user queries the entire database and does not know the document order or page numbers.\par
\noindent * **Alternative**: Use **Section Headers, Table Captions, or Unique Keywords** found in the text.\par
\noindent \hspace*{1em}* Bad: "2페이지에 있는 표를 요약해."\par
\noindent \hspace*{1em}* Good: "'2024년 재무 하이라이트' 표를 요약해."\par

\medskip
\noindent \#\#\#\# Rule 2: Implicit Multi-Page Synthesis\par
\noindent * The query must require information scattered across multiple pages, but **without explicitly stating so**.\par
\noindent * **Strategy**: Identify **Entity A** on one page and **Entity B** on another, then ask about their relationship.\par

\medskip
\noindent \#\#\#\# Rule 3: Entity-Grounded Specificity\par
\noindent * Avoid generic queries like "에너지 정책을 분석해줘."\par
\noindent * Include specific entities found in the text: **Dates, Company Names, Regulations (e.g., ISO-27001), Project Codes, Policy Names, or Program Names**.\par
\noindent * However, do NOT include exact numerical values (see Rule 5).\par

\medskip
\noindent \#\#\#\# Rule 4: Single Natural Query\par
\noindent * The query **MUST be a single unit** appropriate to its format (one question, one instruction, or one keyword cluster).\par
\noindent * **Strictly Forbidden Patterns**:\par
\noindent \hspace*{1em}* Multiple sentences: "\textasciitilde 입니다. \textasciitilde 해주세요."\par
\noindent \hspace*{1em}* Instruction suffixes: "단, \textasciitilde 를 기준으로 답변하시오."\par
\noindent \hspace*{1em}* Explicit output format requests: "\textasciitilde 를 근거로 제시하시오.", "\textasciitilde 를 나열하시오."\par
\noindent \hspace*{1em}* Conditional clauses at the end: "단, \textasciitilde 를 구분하여 제공해야 합니다."\par
\noindent * **Examples**:\par
\noindent \hspace*{1em}* Bad: "2021년과 2022년 상승률을 비교하시오. 단, 수도권과 지방을 구분하여 제시하시오."\par
\noindent \hspace*{1em}* Good: "2021년과 2022년 수도권 및 지방의 주택 매매가격 상승률은 어떻게 달랐나요?"\par

\medskip
\noindent \#\#\#\# Rule 5: Realistic Search Behavior\par
\noindent The query must read as if a **researcher who does NOT have the document** is searching a database by topic and keywords. This single rule covers three aspects:\par

\medskip
\noindent **(a) No Verbatim Document Data**\par
\noindent * Use **conceptual references** (policy names, years, entity names) instead of **exact figures**.\par
\noindent * **Strictly Forbidden**: Specific monetary values, exact percentages, precise statistics copied from the document.\par
\noindent \hspace*{1em}* Bad: "에너지 요금이 €49.5/MWh에서 €94/MWh로 89\% 상승한 이유는?"\par
\noindent \hspace*{1em}* Good: "2022년 프랑스 소매 에너지 요금 급등과 EDF의 ARENH 정책은 어떤 관계가 있나요?"\par

\medskip
\noindent **(b) No Document-Aware Framing**\par
\noindent * **Strictly Forbidden**: "문서에서", "해당 자료의", "위 표에 따르면", "본 보고서의", "제시된 데이터를 기반으로"\par
\noindent * Also forbidden --- **Document Title Scoping** (assumes the user already knows the document exists):\par
\noindent \hspace*{1em}* Bad: "제7차 에너지기본계획에서 원전 비중 목표는?"\par
\noindent \hspace*{1em}* Good: "2025년 일본의 원전 비중 목표"\par

\medskip
\noindent **(c) Realistic User Knowledge**\par
\noindent * The user **knows**: topic area, key entities, time periods of interest.\par
\noindent * The user **does NOT know**: page numbers, document structure, specific numerical values, exact document titles.\par

\medskip
\noindent \#\#\# 4. Query Format Specification\par

\medskip
\noindent \#\#\#\# Question Format (질문형)\par
\noindent * Must be a complete interrogative sentence with question endings.\par
\noindent * **Required elements**: Question word (무엇, 어떻게, 왜, 어떤) OR question ending (\textasciitilde 인가요?, \textasciitilde 있나요?, \textasciitilde 했는가?)\par
\noindent * **Examples**:\par
\noindent \hspace*{1em}* "M2 광의통화 증가율이 2020년 국가채무 증가에 영향을 미쳤는가?"\par
\noindent \hspace*{1em}* "에너지바우처 제도의 지원 대상은 누구인가요?"\par

\medskip
\noindent \#\#\#\# Instruction Format (지시형)\par
\noindent * Must be a command with imperative endings.\par
\noindent * **Required elements**: Imperative ending (\textasciitilde 해주세요, \textasciitilde 하시오, \textasciitilde 분석하라, \textasciitilde 설명하라)\par
\noindent * **Examples**:\par
\noindent \hspace*{1em}* "2020년 M2 통화량과 국가채무 간의 상관관계를 분석해주세요."\par
\noindent \hspace*{1em}* "에너지바우처와 에너지효율개선 사업의 차이점을 비교하라."\par

\medskip
\noindent \#\#\#\# Keyword Format (키워드형)\par
\noindent * **NO complete sentences.** Only noun phrases and search terms.\par
\noindent * **NO verbs, NO question words, NO sentence endings.**\par
\noindent * Mimics search engine input: fragmented, noun-centric.\par

\medskip
\noindent **Keyword Format Rules**:\par
\noindent Allowed / Forbidden\par
\noindent - 명사, 명사구, 복합 명사구 / 동사 (\textasciitilde 하다, \textasciitilde 이다, \textasciitilde 있다)\par
\noindent - 관계 조사 (\textasciitilde 의, \textasciitilde 와/과, \textasciitilde 간, \textasciitilde 에 따른, \textasciitilde 으로 인한) / 질문사 (무엇, 어떻게, 왜, 어떤)\par
\noindent - 고유명사, 연도, 날짜 / 문장 종결 (\textasciitilde 인가요, \textasciitilde 해주세요, \textasciitilde 입니까)\par
\noindent - 관계 표현 (비교, 관계, 영향, 연관성, 상관관계) / 완전한 문장 구조\par
\noindent - 영문 약어 (EDF, ARENH, GDP) / 공백으로만 나열된 독립 키워드들\par
\noindent - 개념적 추상화 표현 / 문서 표 항목명·인덱스의 직접 복사\par

\medskip
\noindent \#\#\#\#\# Keyword Structural Templates\par
\noindent A keyword query must form **one coherent noun phrase**. Every noun must be connected to its neighbors by Korean particles (의, 와/과, 간, 에 따른, 으로 인한, 내, 중, 및) that make the semantic relationship explicit.\par

\medskip
\noindent Templates:\par
\noindent - Comparison: A의 X와/과 B의 Y (간) 차이/비교\par
\noindent \hspace*{1em}Example: "운수업의 부가가치당 에너지소비량과 수송용 에너지소비 비중 차이"\par
\noindent - Correlation: A와/과 B 간 연관성/관계/상관관계\par
\noindent \hspace*{1em}Example: "일반가구의 설계가중치와 도시가구의 에너지소비 간 연관성"\par
\noindent - Causation: A 변화/증가/감소와 B 변화의 연관성/영향\par
\noindent \hspace*{1em}Example: "부산 개별여행 비중 증가와 농수산물 구매 비중 상승의 연관성"\par
\noindent - Condition: A에 따른/으로 인한 B의 변화/추이\par
\noindent \hspace*{1em}Example: "스페인 용량요금 중단에 따른 전력부문 적자의 변화"\par
\noindent - Composition: A 내 B와 C의 비중/분포/구성\par
\noindent \hspace*{1em}Example: "EU 노동 인력 내 녹색 직업과 고도 디지털 집약 직업 간의 연령 분포"\par

\medskip
\noindent **Particle Removal Test**:\par
\noindent Strip all particles (의/와/과/간/에 따른/으로 인한/내/중/및) from the query.\par
\noindent * If the meaning **collapses** -> Well-formed noun phrase.\par
\noindent * If the meaning **stays the same** -> Keyword bag. Rewrite.\par

\medskip
\noindent **Read-Aloud Test**:\par
\noindent Read the query aloud. If there is a natural pause splitting it into two independent chunks with no grammatical bridge -> Two queries glued together. Rewrite.\par

\medskip
\noindent **Bad -> Fixed Examples**:\par
\noindent * "감일도서관 개관 희망도서 바로대출 지역서점 연계 독서문화 활성화 지원 사업 이동도서관 스마트도서관"\par
\noindent \hspace*{1em}-> "감일도서관 개관 이후 희망도서 바로대출 서비스와 지역서점 연계 독서문화 사업 간의 운영 방식 차이"\par
\noindent * "K-방산 폴란드 수출 비중 라틴아메리카 방위비 증가"\par
\noindent \hspace*{1em}-> "K-방산의 폴란드 수출 비중 확대와 라틴아메리카 방위비 증가 간 연관성"\par
\noindent * "베트남 최종 법인세 신고 베트남 개인소득세 체계 동일 과세 기준 여부"\par
\noindent \hspace*{1em}-> "베트남 법인세 최종 신고 체계와 개인소득세 체계의 과세 기준 동일 여부"\par

\medskip
\noindent \#\#\# 5. Quality Checklist (Self-Verification)\par
\noindent Before finalizing, verify ALL checks pass:\par
\noindent - Format Compliance: Query strictly follows the specified format (question/instruction/keyword)\par
\noindent - Single Unit: ONE question, ONE instruction, or ONE keyword phrase -- no multiple sentences\par
\noindent - No Page References: No page numbers, document indices, or positional references\par
\noindent - Realistic Search: No exact values from the document, no document-aware framing, no document title scoping (Rule 5)\par
\noindent - Entity-Grounded: Includes searchable entities (names, years, policy names) but not verbatim data\par
\noindent - Multi-Page Implicit: Requires information from multiple pages without explicitly stating it\par
\noindent - Keyword Coherence (keyword only): **Particle Removal Test** passes: stripping particles must break the meaning\par
\noindent - Single Phrase (keyword only): **Read-Aloud Test** passes: query flows as one utterance with no independent chunks\par

\medskip
\noindent \#\#\# 6. Output Generation\par
\noindent Generate the output strictly adhering to the defined JSON schema.\par
\noindent The query must be in **Korean** and must pass all checks in the Quality Checklist above.\par
\noindent **Pay special attention to the Query Format specification---the linguistic structure must match exactly.**\par

\end{tcolorbox}
\captionsetup{hypcap=false}
\captionof{figure}{Prompt for generating context-based retrieval queries with controls for realism, diversity, and query formulation}
\label{fig:query-from-context-prompt}
\end{center}

\twocolumn

\begin{figure*}[t]
    \centering
    \includegraphics[width=\textwidth]{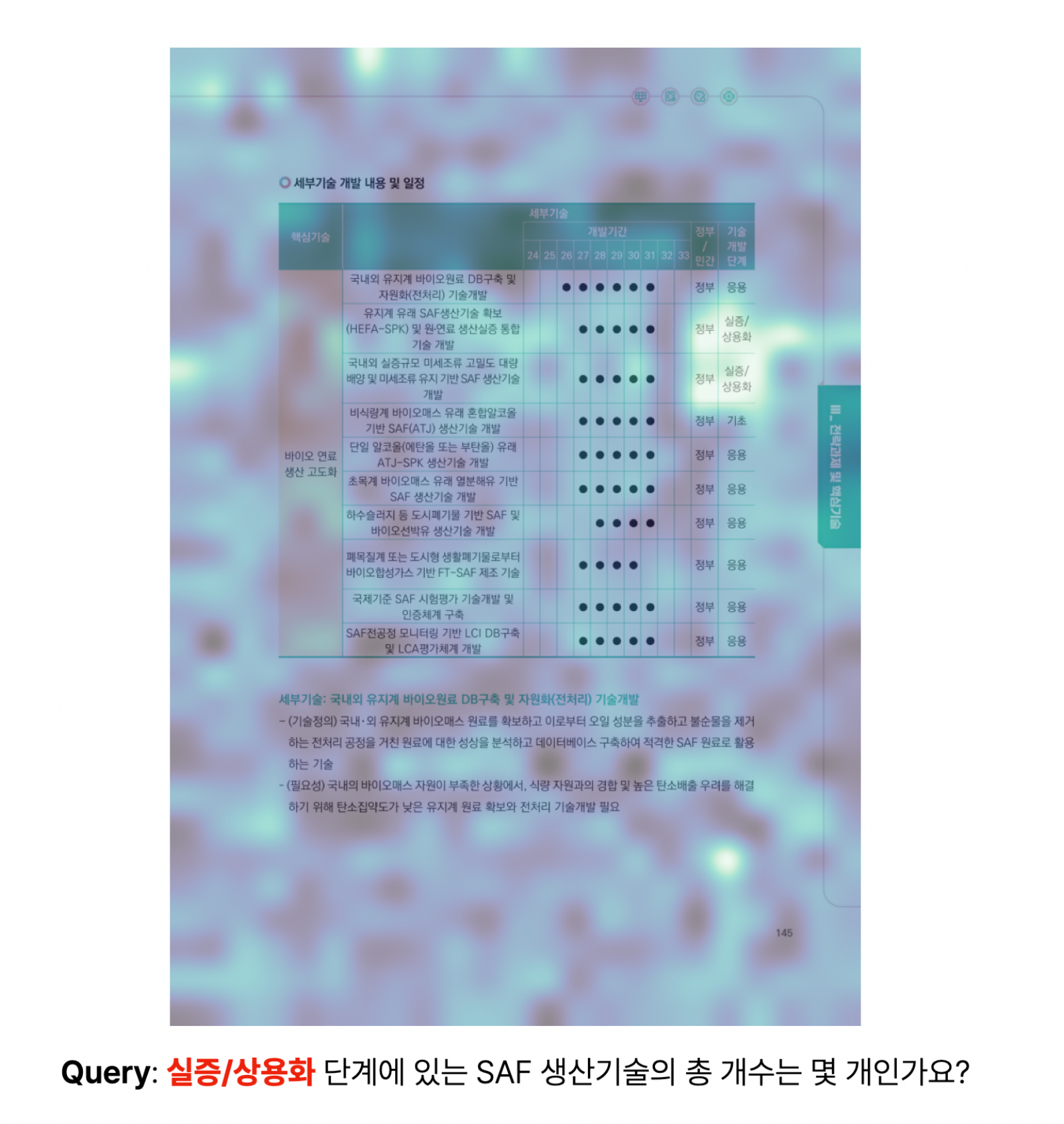}
    \caption{
    Additional query-to-document similarity heatmap for the fine-tuned \texttt{colqwen2-v1.0} model on a Korean document example. The model assigns high similarity to image patches corresponding to the term ``실증/상용화,'' highlighting its focus on query-relevant textual evidence.
    }
    \label{fig:interpretability_appendix}
\end{figure*}

\begin{table*}[t]
\centering
\scriptsize
\setlength{\tabcolsep}{4pt}
\renewcommand{\arraystretch}{1.1}
\begin{tabular}{p{8.6cm}p{2.6cm}c p{2.0cm}}
\toprule
\textbf{Title} & \textbf{Provider} & \textbf{Pages} & \textbf{License} \\
\midrule

\rowcolor{gray!15}
\multicolumn{4}{c}{\textit{Cybersecurity}} \\
\midrule
\href{https://www.data.go.kr/data/15102015/fileData.do}{갠드크랩 랜섬웨어 악성코드 분석 기술 보고서}        & 한국인터넷진흥원 & 92  & No Restriction \\
\href{https://www.data.go.kr/data/15102009/fileData.do}{사이버위협 동향보고서 (Windows 취약점 동향 및 업데이트 정책 등)}              & 한국인터넷진흥원 & 104 & No Restriction \\
\href{https://www.data.go.kr/data/15102010/fileData.do}{사이버위협 동향보고서 (동형암호 기반 데이터 결합 및 분석 등)}   & 한국인터넷진흥원 & 100 & No Restriction \\
\href{https://www.data.go.kr/data/15066710/fileData.do}{사이버 위협 동향보고서 (피싱 메일 공격 사례 등)}        & 한국인터넷진흥원 & 96  & No Restriction \\
\href{https://www.data.go.kr/data/15066723/fileData.do}{사이버 위협 동향보고서 (기업 보안관리자의 크리덴셜 스터핑(Credential Stuffing) 공격 대응방안 등)}       & 한국인터넷진흥원 & 104 & No Restriction \\
\href{https://www.data.go.kr/data/15102011/fileData.do}{사이버위협 동향보고서 (공인인증서 문제점과 DID 기술 전망 등)}      & 한국인터넷진흥원 & 100 & No Restriction \\
\href{https://www.data.go.kr/data/15102014/fileData.do}{사이버위협 동향보고서 (ATT\&CK Framework 개념과 이해 등)}           & 한국인터넷진흥원 & 80  & No Restriction \\
\href{https://www.data.go.kr/data/15102017/fileData.do}{랜섬웨어 대응을 위한 안전한 정보시스템 백업 가이드(개정본)}  & 한국인터넷진흥원 & 68  & No Restriction \\
\href{https://www.data.go.kr/data/15156664/fileData.do}{해킹진단도구 활용 사례 (취약한 관리자 계정 악용을 악용한 데이터 유출)}    & 한국인터넷진흥원 & 23  & No Restriction \\
\href{https://www.data.go.kr/data/15156665/fileData.do}{해킹진단도구 활용 사례 (노출된 SMB 파일 서버를 통한 AD 환경 장악)}       & 한국인터넷진흥원 & 23  & No Restriction \\
\href{https://www.data.go.kr/data/15138872/fileData.do}{해킹진단도구 활용 사례 (취약한 MS-SQL 서버를 통한 랜섬웨어 침해사고 )}         & 한국인터넷진흥원 & 14  & No Restriction \\
\href{https://www.data.go.kr/data/15139554/fileData.do}{해킹진단도구 활용 사례 (AD 환경에서의 RAT 악성코드 감염)} & 한국인터넷진흥원 & 16  & No Restriction \\
\href{https://www.data.go.kr/data/15102022/fileData.do}{AD서버 악용 내부망 랜섬웨어 유포 사례 분석}          & 한국인터넷진흥원 & 31  & No Restriction \\
\href{https://www.data.go.kr/data/15102027/fileData.do}{Log4j 위협 대응 보고서}                              & 한국인터넷진흥원 & 35  & No Restriction \\
\href{https://www.data.go.kr/data/15138752/fileData.do}{NAS 보안 가이드}                                     & 한국인터넷진흥원 & 161 & No Restriction \\
\href{https://www.data.go.kr/data/15100322/fileData.do}{TTPs2 스피어 피싱을 통한 공격망 구성 방식 분석}     & 한국인터넷진흥원 & 79  & No Restriction \\
\href{https://www.data.go.kr/data/15100324/fileData.do}{TTPs3 공격자의 악성코드 활용 전략 분석}             & 한국인터넷진흥원 & 27  & No Restriction \\

\midrule

\rowcolor{gray!15}
\multicolumn{4}{c}{\textit{Economic}} \\
\midrule
\href{https://www.data.go.kr/data/15066779/fileData.do}{최근 경제동향 (2021. 3월호)} & 기획재정부 & 80 & No Restriction \\
\href{https://www.data.go.kr/data/15066779/fileData.do}{최근 경제동향 (2021. 6월호)} & 기획재정부 & 80 & No Restriction \\
\href{https://www.data.go.kr/data/15066779/fileData.do}{최근 경제동향 (2021. 9월호)} & 기획재정부 & 80 & No Restriction \\
\href{https://www.data.go.kr/data/15066779/fileData.do}{최근 경제동향 (2021. 12월호)} & 기획재정부 & 80 & No Restriction \\
\href{https://www.data.go.kr/data/15066779/fileData.do}{최근 경제동향 (2022. 3월호)} & 기획재정부 & 80 & No Restriction \\
\href{https://www.data.go.kr/data/15066779/fileData.do}{최근 경제동향 (2022. 6월호)} & 기획재정부 & 80 & No Restriction \\
\href{https://www.data.go.kr/data/15066779/fileData.do}{최근 경제동향 (2022. 9월호)} & 기획재정부 & 80 & No Restriction \\
\href{https://www.data.go.kr/data/15066779/fileData.do}{최근 경제동향 (2022. 12월호)} & 기획재정부 & 80 & No Restriction \\
\href{https://www.data.go.kr/data/15066779/fileData.do}{최근 경제동향 (2023. 3월호)} & 기획재정부 & 82 & No Restriction \\
\href{https://www.data.go.kr/data/15066779/fileData.do}{최근 경제동향 (2023. 6월호)} & 기획재정부 & 80 & No Restriction \\
\href{https://www.data.go.kr/data/15066779/fileData.do}{최근 경제동향 (2023. 9월호)} & 기획재정부 & 80 & No Restriction \\
\href{https://www.data.go.kr/data/15066779/fileData.do}{최근 경제동향 (2023. 12월호)} & 기획재정부 & 80 & No Restriction \\
\href{https://www.data.go.kr/data/15066779/fileData.do}{최근 경제동향 (2024. 3월호)} & 기획재정부 & 77 & No Restriction \\
\href{https://www.data.go.kr/data/15066779/fileData.do}{최근 경제동향 (2024. 6월호)} & 기획재정부 & 77 & No Restriction \\
\href{https://www.data.go.kr/data/15066779/fileData.do}{최근 경제동향 (2024. 9월호)} & 기획재정부 & 77 & No Restriction \\
\href{https://www.data.go.kr/data/15066779/fileData.do}{최근 경제동향 (2024. 12월호)} & 기획재정부 & 77 & No Restriction \\
\href{https://www.data.go.kr/data/15066779/fileData.do}{최근 경제동향 (2025. 3월호)} & 기획재정부 & 77 & No Restriction \\
\href{https://www.data.go.kr/data/15066779/fileData.do}{최근 경제동향 (2025. 6월호)} & 기획재정부 & 77 & No Restriction \\
\href{https://www.data.go.kr/data/15066779/fileData.do}{최근 경제동향 (2025. 9월호)} & 기획재정부 & 77 & No Restriction \\
\href{https://www.data.go.kr/data/15066779/fileData.do}{최근 경제동향 (2025. 12월호)} & 기획재정부 & 77 & No Restriction \\

\midrule

\rowcolor{gray!15}
\multicolumn{4}{c}{\textit{Energy}} \\
\midrule
\href{https://www.data.go.kr/data/15095909/fileData.do}{해외전력산업동향 (2017 China)} & 한국전력거래소 & 57 & No Restriction \\
\href{https://www.data.go.kr/data/15095909/fileData.do}{해외전력산업동향 (2017 Japan)} & 한국전력거래소 & 43 & No Restriction \\
\href{https://www.data.go.kr/data/15095909/fileData.do}{해외전력산업동향 (2017 USA)} & 한국전력거래소 & 43 & No Restriction \\
\href{https://data.etic.kr/brd/m_11/view.do?seq=34}{제4차 에너지기술개발계획 기술로드맵: 에너지저장} & 한국에너지기술평가원 & 172 & KOGL Type 2 \\
\href{https://data.etic.kr/brd/m_11/view.do?seq=34}{제4차 에너지기술개발계획 기술로드맵: 총괄} & 한국에너지기술평가원 & 76 & KOGL Type 2 \\
\href{https://www.data.go.kr/data/15081561/fileData.do}{대전광역시 신재생에너지 보급계획} & 대전광역시 & 536 & No Restriction \\
\href{https://www.data.go.kr/data/15095909/fileData.do}{해외전력산업동향 (2023)} & 한국전력거래소 & 478 & No Restriction \\
\href{https://www.data.go.kr/data/15055169/fileData.do}{인천광역시 에너지백서} & 인천광역시 & 259 & No Restriction \\
\href{https://data.etic.kr/brd/m_11/view.do?seq=34}{제5차 에너지기술개발계획 기술로드맵: 수요관리} & 한국에너지기술평가원 & 85 & KOGL Type 2 \\
\href{https://data.etic.kr/brd/m_11/view.do?seq=34}{제5차 에너지기술개발계획 기술로드맵: 효율향상} & 한국에너지기술평가원 & 162 & KOGL Type 2 \\
\href{https://www.data.go.kr/data/15007794/fileData.do}{에너지 기술정책 포커스 (2025 주요국 기후에너지정책)} & 한국에너지기술연구원 & 122 & No Restriction \\

\midrule

\rowcolor{gray!15}
\multicolumn{4}{c}{\textit{HR}} \\
\midrule
\href{https://www.data.go.kr/data/3038238/fileData.do}{고용형태별 근로실태조사 보고서} & 고용노동부 & 277 & KOGL Type 1 \\
\href{https://www.data.go.kr/data/15033110/fileData.do}{블라인드 채용 가이드북} & 고용노동부 & 88 & No Restriction \\
\href{https://www.data.go.kr/data/15032836/fileData.do}{일·가정 양립 실태조사 보고서} & 고용노동부 & 423 & No Restriction \\
\href{https://www.data.go.kr/data/15140285/fileData.do}{한국직업전망} & 한국고용정보원 & 668 & KOGL Type 2 \\
\href{https://www.data.go.kr/data/15145184/fileData.do}{유망 신산업 산업기술인력 전망: 이차전지} & 한국산업기술진흥원 & 134 & No Restriction \\
\href{https://www.data.go.kr/data/15145184/fileData.do}{유망 신산업 산업기술인력 전망: 첨단화학소재} & 한국산업기술진흥원 & 134 & No Restriction \\
\href{https://www.data.go.kr/data/15145184/fileData.do}{유망 신산업 산업기술인력 전망: 첨단섬유소재} & 한국산업기술진흥원 & 148 & No Restriction \\
\href{https://www.data.go.kr/data/15145184/fileData.do}{유망 신산업 산업기술인력 전망: 신금속소재} & 한국산업기술진흥원 & 136 & No Restriction \\
\href{https://www.data.go.kr/data/15145184/fileData.do}{유망 신산업 산업기술인력 전망: 차세대세라믹소재} & 한국산업기술진흥원 & 138 & No Restriction \\

\bottomrule
\end{tabular}
\caption{Document metadata for KoViDoRe across all subsets.}
\label{tab:kovidore_metadata}
\end{table*}

\begin{table*}[t]
\centering
\scriptsize
\setlength{\tabcolsep}{4pt}
\renewcommand{\arraystretch}{1.03}
\begin{tabular}{p{8.6cm}p{2.6cm}c p{2.0cm}}
\toprule
\textbf{Title} & \textbf{Provider} & \textbf{Pages} & \textbf{License} \\
\midrule
\href{https://www.data.go.kr/data/15157109/fileData.do}{에너지총조사} & 기후에너지환경부 & 774 & No Restrictions \\
\href{https://www.data.go.kr/data/15107714/fileData.do}{주요업무계획} & 경기도 하남시 & 640 & No Restrictions \\
\href{https://www.data.go.kr/data/15038527/fileData.do}{지방공무원 인사실무} & 행정안전부 & 521 & No Restrictions \\
\href{https://www.data.go.kr/data/15088366/fileData.do}{항만편람} & 해양수산부 & 515 & No Restrictions \\
\href{https://www.data.go.kr/data/15090298/fileData.do}{국가연구개발사업 상위평가보고서} & 과학기술정보통신부 & 479 & No Restrictions \\
\href{https://www.data.go.kr/data/15154055/fileData.do}{연구보고서 현황} & 한국방송통신전파진흥원 & 425 & No Restrictions \\
\href{https://www.data.go.kr/data/15117879/fileData.do}{작업환경실태조사 보고서} & 한국산업안전보건공단 & 363 & No Restrictions \\
\href{https://www.data.go.kr/data/3071773/fileData.do}{국가연구개발사업 특정평가보고서} & 과학기술정보통신부 & 323 & No Restrictions \\
\href{https://www.data.go.kr/data/15089055/fileData.do}{관리형매립지 조사결과보고서} & 수도권매립지관리공사 & 285 & No Restrictions \\
\href{https://www.data.go.kr/data/15145080/fileData.do}{ICT 융복합 시설의 안전한 전자파 환경 기반 조성 연구} & 국립전파연구원 & 253 & KOGL Type 1 \\
\href{https://www.data.go.kr/data/15145081/fileData.do}{전자파 흡수전력밀도 등 전자파 인체노출량 평가기술 연구} & 국립전파연구원 & 249 & KOGL Type 1 \\
\href{https://www.data.go.kr/data/15156787/fileData.do}{해외건설 세무업무 매뉴얼} & 국토교통부 & 216 & No Restrictions \\
\href{https://www.data.go.kr/data/15156699/fileData.do}{처분시설 부지주변 방사선환경조사 보고서} & 한국원자력환경공단 & 211 & KOGL Type 1 \\
\href{https://www.data.go.kr/data/15156970/fileData.do}{스마트 안전유지관리 시설물 확대방안 마련 용역 보고서} & 국토안전관리원 & 210 & No Restrictions \\
\href{https://www.data.go.kr/data/15075905/fileData.do}{해양수산발전기본계획} & 해양수산부 & 204 & No Restrictions \\
\href{https://www.data.go.kr/data/15117941/fileData.do}{디지털미디어 허브 조성을 위한 빛마루 중장기 전략 연구보고서} & 한국방송통신전파진흥원 & 195 & No Restrictions \\
\href{https://www.data.go.kr/data/3038031/fileData.do}{유엔개황 책자} & 외교부 & 178 & No Restrictions \\
\href{https://www.data.go.kr/data/3038240/fileData.do}{기업체노동비용조사 보고서} & 고용노동부 & 153 & No Restrictions \\
\href{https://www.data.go.kr/data/15063835/fileData.do}{(PDF)인삼재배전서} & 경상북도 & 151 & No Restrictions \\
\href{https://www.data.go.kr/data/15117938/fileData.do}{디지털미디어 신산업 진흥 방안 및 인력수급 기초조사에 관한 연구보고서} & 한국방송통신전파진흥원 & 146 & No Restrictions \\
\href{https://www.data.go.kr/data/15112219/fileData.do}{무인도서 100선} & 해양수산부 & 125 & KOGL Type 1 \\
\href{https://www.data.go.kr/data/15076006/fileData.do}{환경관리해역 기본계획} & 해양수산부 & 125 & No Restrictions \\
\href{https://www.data.go.kr/data/15156776/fileData.do}{해외건설 법률컨설팅 사례} & 국토교통부 & 121 & No Restrictions \\
\href{https://www.data.go.kr/data/15117937/fileData.do}{국내외 온라인 동영상 미디어 콘텐츠 시장 전망 및 정책 추진방향 연구보고서} & 한국방송통신전파진흥원 & 115 & No Restrictions \\
\href{https://www.data.go.kr/data/15142321/fileData.do}{합성데이터 생성 활용 안내서} & 개인정보보호위원회 & 110 & KOGL Type 1 \\
\href{https://www.data.go.kr/data/15076006/fileData.do}{환경관리해역 기본계획} & 해양수산부 & 90 & No Restrictions \\
\href{https://www.data.go.kr/data/15105424/fileData.do}{국가공무원인재개발원 교육운영계획} & 인사혁신처 & 85 & No Restrictions \\
\href{https://www.data.go.kr/data/15024091/fileData.do}{중소기업 경제동향 정보} & 중소벤처기업연구원 & 75 & No Restrictions \\
\href{https://www.data.go.kr/data/15142329/fileData.do}{생체정보 보호 안내서} & 개인정보보호위원회 & 73 & KOGL Type 1 \\
\href{https://www.data.go.kr/data/15061586/fileData.do}{인재개발 종합계획} & 인사혁신처 & 68 & No Restrictions \\
\href{https://www.data.go.kr/data/15040380/fileData.do}{공공외교 기본계획} & 외교부 & 59 & No Restrictions \\
\href{https://www.data.go.kr/data/3081488/fileData.do}{관광실태조사 정보} & 부산광역시 & 58 & No Restrictions \\
\href{https://www.data.go.kr/data/15044488/fileData.do}{카지노 비즈니스와 제도} & 그랜드코리아레저(주) & 57 & No Restrictions \\
\href{https://www.data.go.kr/data/15039602/fileData.do}{모바일 전자정부서비스 앱 소스코드 검증 가이드라인} & 행정안전부 & 51 & No Restrictions \\
\href{https://www.data.go.kr/data/15142243/fileData.do}{개인정보 유출 등 사고 대응 매뉴얼} & 개인정보보호위원회 & 48 & KOGL Type 1 \\
\href{https://www.data.go.kr/data/15153323/fileData.do}{교통 리포트} & 서울특별시 & 42 & No Restrictions \\
\href{https://www.data.go.kr/data/15077372/fileData.do}{블랙잭 게임의 이해} & 그랜드코리아레저(주) & 32 & No Restrictions \\
\href{https://www.data.go.kr/data/15145110/fileData.do}{개인정보 유출 신고 동향 및 예방 방법} & 한국인터넷진흥원 & 32 & No Restrictions \\
\href{https://www.data.go.kr/data/15156813/fileData.do}{i SMR 및 SSNC 설명자료} & 한국수력원자력(주) & 27 & No Restrictions \\
\href{https://www.data.go.kr/data/15150187/fileData.do}{농지개량행위 신고 업무지침} & 농림축산식품부 & 24 & No Restrictions \\
\href{https://www.data.go.kr/data/15112345/fileData.do}{미디어이슈\_광고요금제 도입을 앞둔 넷플릭스에 대한 인식 및 이용 조사} & 한국언론진흥재단 & 23 & KOGL Type 1 \\
\href{https://www.data.go.kr/data/15090118/fileData.do}{위성전파 감시 정보} & 중앙전파관리소 & 19 & KOGL Type 1 \\
\href{https://www.data.go.kr/data/15112343/fileData.do}{미디어이슈\_이대남 현상에 대한 인식} & 한국언론진흥재단 & 19 & KOGL Type 1 \\
\href{https://www.data.go.kr/data/15086396/fileData.do}{미디어이슈\_코로나19 관련 정보 이용 및 인식 현황} & 한국언론진흥재단 & 19 & KOGL Type 1 \\
\href{https://www.data.go.kr/data/15151306/fileData.do}{해외시장 신용위험 보고서} & 한국무역보험공사 & 18 & No Restrictions \\
\href{https://www.data.go.kr/data/15049659/fileData.do}{중남미 관련 보고서 (제약바이오)} & 외교부 & 12 & No Restrictions \\
\href{https://www.data.go.kr/data/15049659/fileData.do}{중남미 관련 보고서 (방위산업)} & 외교부 & 10 & No Restrictions \\
\href{https://www.data.go.kr/data/15156629/fileData.do}{노인 일자리 및 사회활동 지원사업 시행 20년의 성과와 발전과제} & 한국노인인력개발원 & 9 & No Restrictions \\
\href{https://www.data.go.kr/data/15117914/fileData.do}{전국 주배관망 가스인입지점별 인입가능량} & 한국가스공사 & 3 & No Restrictions \\
\bottomrule
\end{tabular}
\caption{Document metadata for Ko-VDR Train Public.}
\label{tab:ko_vdr_train_public_metadata}
\end{table*}

\end{document}